\documentclass[pdflatex,sn-mathphys-num]{sn-jnl}

\usepackage{graphicx}%
\usepackage{multirow}%
\usepackage{amsmath,amssymb,amsfonts}%
\IfFileExists{dsfont.sty}{\usepackage{dsfont}\newcommand{\indicator}{\mathds{1}}}{\newcommand{\indicator}{\mathbf{1}}}%
\usepackage{amsthm}%
\usepackage{mathrsfs}%
\usepackage[title]{appendix}%
\usepackage{xcolor}%
\usepackage{textcomp}%
\usepackage{manyfoot}%
\usepackage{booktabs}%
\usepackage{array}%
\usepackage{algorithm}%
\usepackage{algorithmicx}%
\usepackage{algpseudocode}%
\usepackage{listings}%
\usepackage[section,below]{placeins}
\usepackage[all]{hypcap}
\hypersetup{bookmarksdepth=subsubsection}
\makeatletter\@ifpackageloaded{geometry}{\geometry{hcentering,bindingoffset=0mm}}{}\makeatother
\theoremstyle{thmstyleone}%
\theoremstyle{thmstyletwo}%

\theoremstyle{thmstylethree}%

\begin{document}

\title[Beyond Marginal Monitoring]{Beyond Marginal Monitoring: Distributed Joint-Distribution Testing for Data Concept Drift in Large Scale E-Commerce Operations}


\author*[1]{\fnm{\c{C}a\u{g}da\c{s}} \sur{Pullu}}\email{cagdas.pullu@trendyol.com}

\author[1]{\fnm{Mahmut Emir} \sur{Arslan}}\email{emir.arslan@trendyol.com}

\author[1]{\fnm{Bu\u{g}ra} \sur{Balka\c{c}}}\email{bugra.balkac@trendyol.com}

\author[1]{\fnm{Aylin \"{O}ndersev} \sur{Balta}}\email{aylin.ondersev@trendyol.com}

\author[1]{\fnm{Cihangir Celal} \sur{Palac\i{}}}\email{cihangir.palaci@trendyol.com}

\author[1]{\fnm{Fikri Cem} \sur{Y\i{}lmaz}}\email{fikricem.yilmaz@trendyol.com}

\author[2]{\fnm{Altan} \sur{Cakir}}\email{cakirmua@itu.edu.tr}

\affil*[1]{\orgdiv{Data \& Analytics}, \orgname{Trendyol Group}, \orgaddress{\city{Istanbul}, \postcode{34485}, \state{Istanbul}, \country{T\"{u}rkiye}}}

\affil[2]{\orgdiv{Department of Data Science and Analytics}, \orgname{Istanbul Technical University}, \orgaddress{\street{Maslak}, \city{Istanbul}, \postcode{34469}, \state{Istanbul}, \country{T\"{u}rkiye}}}


\abstract{Concept drift threatens production machine learning, yet the empirical behavior of multivariate two-sample drift detectors at scale remains under-characterized. Existing benchmarks rarely address the hundreds of millions of rows and high-cardinality features typical of industrial-operational datasets. We evaluate five multi-column two-sample tests (marginal, projection-based, and kernel embedding methods) across three complementary environments: the Harvard Dataverse, a validated Failing Loudly reproduction (mean absolute error \(\in [0.030, 0.053]\)), and a novel synthetic-injection benchmark on the $137.5$-million-row Trendyol collection-ranking feature table. Testing four drift types across two severity--scope regimes, we demonstrate that distributed Maximum Mean Discrepancy with Random Fourier Features on Apache Spark scales robustly. Averaged over the four drift types in the strong regime and under a calibrated threshold, it achieves a Pearson correlation of \(r = 0.940\) with expected drift magnitude, an \(80.4\%\) true positive rate, and a \(3.2\%\) false positive rate. Conversely, the per-dimension Kolmogorov--Smirnov test failed due to statistic saturation from ID-like columns under asymmetric sampling, establishing a critical constraint for large-scale sampling design. At weak configurations (realized-flip fractions \(\le 0.57\%\)), detectors struggled to reliably discriminate, highlighting the need for future intensity-grid power analyses to distinguish fundamental sensitivity bounds from scalable threshold shifts.}


\maketitle

\section{Introduction}\label{sec1}

The integration of machine learning models into live products means that the data encountered in production may drift away from the training distribution. This \emph{dataset shift} erodes prediction accuracy and, in the absence of continuous human oversight, progresses silently \citep{morenotorres2012unifying}. Where label-oriented evaluation feedback arrives slowly or not at all---as with sparse post-order return notifications or moderator approvals---drift is detected with two-sample tests on the data available at deployment: the features \(X\) and, when an outcome or prediction channel exists, the label variable \(Y\) produced by (or emitted alongside) the model. These tests are conventionally termed \emph{unsupervised}, yet \emph{what the test observes}---\(X\) alone or the joint vector \((X, Y)\)---must be delineated carefully. At Trendyol\footnote{Founded in 2010, Trendyol is Turkiye’s leading e-commerce ecosystem, integrating core e-commerce, smart logistics, and financial technologies. Operating in 36 countries across Europe and the Gulf Region, its extensive technological infrastructure and high-volume data streams provide a robust foundation for big data and AI-based research and development projects.}, a regional leader in e-commerce, the production table lacks a genuine outcome column, so we construct a synthetic benchmark target \(Y_{\text{synth}}\) from live-variance covariates (Section~\ref{subsec5-4}); it is neither an observed outcome nor a model prediction, and the results must be interpreted with this distinction in mind.

Two research fields have addressed this problem independently. The stream-based concept drift community formulates regimes in which \(P(Y \mid X)\) changes while \(P(X)\) stays stationary \citep{gama2014survey,lopez2020streams}, and the multivariate two-sample testing community measures the detection accuracy of per-dimension and kernel tests under controlled covariate shifts \citep{gretton2012kernel,rabanser2019failing}. Both rely on synthetic protocols, and the behavior of these tests at \emph{production scale}---across millions of rows and wide tables with hundreds of columns---remains unresolved. D3Bench \citep{muller2024d3bench}, the framework of Cerqueira et al.\ \citep{cerqueira2026framework}, and the controlled comparison of unsupervised detectors by Hinder et al.\ \citep{hinder2024survey} approach this gap but remain limited to synthetic or small-scale scenarios, while learner-based approaches \citep{khan2026learner} and conformal calibration proposals \citep{leong2026conformal} address production settings without labeled evaluation feedback from a different angle.

We validate two-sample tests across three complementary layers. Batch and sliding-scan protocols on the Harvard Dataverse streams test whether drift is detected and localized; a numerical reproduction of the Failing Loudly benchmark verifies that our implementations match the published results; and a synthetic concept drift injection into the \({\sim}137.5\)-million-row Trendyol collection-ranking feature table, which feeds Trendyol's live recommendation system, tests behavior at production scale with eight scenarios (four drift types \(\times\) two severity--scope regimes) and five multi-column tests on an 11-dimensional benchmark vector. In every layer the tests compare the joint vector \((X, Y)\), with the generator's class label in the Harvard Dataverse streams and \(Y_{\text{synth}}\) in the Trendyol benchmark. They are unsupervised in that they need no labeled evaluation feedback, but they are not \(X\)-only: even when the injection leaves \(X\) unchanged row by row, they can capture the change in the \(Y\) marginal (Sections~\ref{subsec5-4} and~\ref{subsec6-3}).

Our primary findings are four. 
(i) Distributed Spark Maximum Mean Discrepancy with Random Fourier Features (RFF-MMD) is the most consistent test at production scale: in the strong regime and under a null-calibrated threshold, it reaches, averaged over the four drift types, a Pearson correlation of \(r = 0.940\) with the per-bucket expected drift magnitude, a true positive rate (TPR) of \(80.4\%\), and a false positive rate (FPR) of \(3.2\%\) (Section~\ref{subsec6-3}). Spark Sliced Wasserstein Distance (SWD) tracks drift comparably (\(r = 0.819\)) but at an FPR of about \(10\%\) under its fixed threshold. 
(ii) Bonferroni-aggregated per-dimension Kolmogorov--Smirnov (KS) fails not through the classical \(n \to \infty\) \(p\)-value collapse but through statistic saturation by ID-like columns under the asymmetric reference/current sampling design (\(r = -0.085\), FPR \(=\) TPR \(= 100\%\); Section~\ref{sec7}). 
(iii) At the weak configuration (realized-flip fractions of \(0.29\)--\(0.57\%\)), neither RFF-MMD nor SWD discriminates reliably, which marks an empirical low-signal failure point for the present setting. 
(iv) A source-table pre-check and a four-part sanity-check protocol (Appendix~\ref{app-sanity}) form a recommended workflow for production drift benchmarks; without them, benchmark results risk measuring incidental data-quality artifacts rather than the intended signal.

Section~\ref{sec2} reviews related work, Sections~\ref{sec3}--\ref{sec5} present the methodology and experimental design, Section~\ref{sec6} reports the results, Section~\ref{sec7} discusses them, and Section~\ref{sec8} concludes.

\section{Related Work}\label{sec2}

\begin{figure}[!htbp]
\centering
\includegraphics[width=0.8\textwidth]{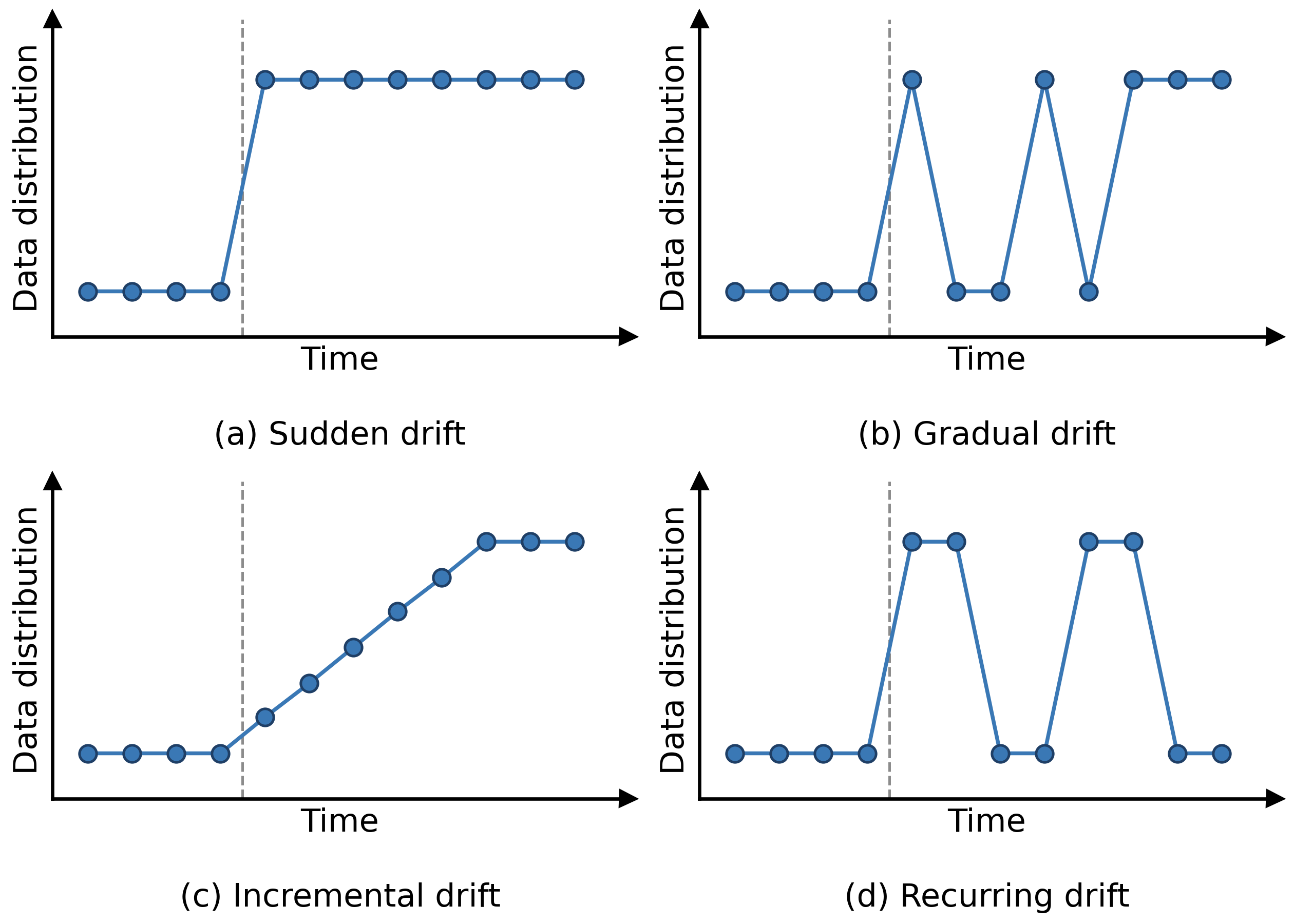}
\caption{Temporal concept-drift profiles, following the taxonomy of \citet{gama2014survey}: (a) sudden drift replaces the old concept at once; (b) gradual drift alternates between the old and new concepts until the new one prevails; (c) incremental drift moves through intermediate states; (d) recurring drift makes previously seen concepts reappear. The dashed line marks the drift onset. Terminology varies across surveys; for example, \citet{hinder2024survey} describe incremental drift as sampling from both concepts with varying probabilities.}\label{fig:drift-taxonomy}
\end{figure}

\citet{gama2014survey} give the canonical taxonomy of distributional shift and the four temporal drift profiles (sudden, gradual, incremental, recurring; Figure~\ref{fig:drift-taxonomy}), with detection lineages including error-monitoring \citep{gama2004learning}, adaptive windows \citep{bifet2007learning}, and concept-history/meta-learning for recurring shifts \citep{suarezcetrulo2023survey}. The \emph{Failing Loudly} benchmark of \citet{rabanser2019failing} evaluates unsupervised shift detection on image test sets (MNIST, CIFAR-10) by combining dimensionality reduction with two-sample testing: multiple univariate Kolmogorov--Smirnov tests aggregated by the Bonferroni correction (on raw features and on sparse random projections, SRP), kernel Maximum Mean Discrepancy with a radial basis function (RBF) kernel (MMD-RBF), which compares the mean embeddings of the two distributions in a reproducing kernel Hilbert space (RKHS) and, under a characteristic kernel, responds to any difference in the full joint distribution \citep{gretton2012kernel}, a chi-square test on hard predicted labels, and a binomial test on a domain classifier; we reproduce the detection accuracies of its KS and MMD variants as validation (Section~\ref{subsec6-2}). The univariate marginal battery used in our Harvard Dataverse layer (chi-square, Jensen--Shannon, Kullback--Leibler, Population Stability Index, Kolmogorov--Smirnov, Wasserstein; Table~\ref{tab:tests-overview}) is our own addition. The \emph{Harvard Dataverse} concept-drift streams of \citet{lopez2020streams} provide five stream families (SEA, SINE, STAGGER, MIXED, RT) with stationary \(P(X)\) by construction, isolating \(P(Y \mid X)\) shifts and providing a controlled testbed for distinguishing multi-column from univariate marginal tests (Section~\ref{subsec6-1}). Other real-world benchmarks include the Insects benchmark \citep{souza2020challenges} (\({\sim}4.7\)M rows, physical sensor data). We use the Harvard Dataverse and Failing Loudly benchmarks as validation layers before the production-scale synthetic drift injection on the Trendyol collection-ranking feature table (Sections~\ref{subsec5-4}, \ref{subsec6-3}). The distributed statistical computations are executed on Apache Spark \citep{zaharia2012rdds} and its SQL/DataFrame layer \citep{armbrust2015sparksql}; cluster configuration details in Section~\ref{subsec4-2}. Table~\ref{tab:tests-overview} summarizes the two-sample tests evaluated across the three layers along with their families, decision rules, and complexity properties.

\begin{table}[!htbp]
\caption{Overview of two-sample tests evaluated in this study. Family classification follows Section~\ref{subsec3-1} (marginal per-feature; marginal per-feature aggregation on the stacked vector; projection-based; multivariate embedding/kernel). Vector column indicates the input to which the test is applied. Complexity is per-position at reference size \(n_R\) and current-window size \(n_C\). Parameter values are those of the Failing Loudly reproduction; the Trendyol scan settings are given in Section~\ref{subsubsec6-3-1}.}\label{tab:tests-overview}
\centering
\footnotesize
\begin{tabular}{@{}>{\raggedright\arraybackslash}p{0.19\textwidth}>{\raggedright\arraybackslash}p{0.16\textwidth}>{\raggedright\arraybackslash}p{0.16\textwidth}>{\raggedright\arraybackslash}p{0.18\textwidth}>{\raggedright\arraybackslash}p{0.17\textwidth}@{}}
\toprule
\textbf{Test} & \textbf{Family} & \textbf{Vector} & \textbf{Decision Rule / Calibration} & \textbf{Complexity} \\ \midrule
\textbf{Univariate marginal \(\times 6\)}\footnotemark[1] & Marginal per-feature & \(X_k\) & Fixed threshold or asymptotic \(p\) & \(O(n)\) per column \\
\textbf{Per-dim KS} & Marginal aggregation & \((X, Y)\) stack & Bonferroni: \(\min_k p_k < \alpha_0/d\) & \(O(nd \log n)\) \\
\textbf{Projected KS (SRP+KS)} & Projection-based & \((X, Y) S\), \(S \in \mathbb{R}^{d \times 32}\) & Bonferroni on projected dims & \(O(nd + n \log n)\) \\
\textbf{MMD-RBF} & Multivariate kernel & \((X, Y)\) stack & Permutation (\(B = 1{,}000\)) & \(O(n^2 d)\); kernel matrix \\
\textbf{RFF-MMD} & Multivariate kernel & \((X, Y)\) stack & Permutation (\(B = 200\)) & \(O(n D d)\); \(D = 500\) \\
\textbf{Sliced Wasserstein Distance} & Projection-based & \((X, Y)\) projections & Fixed \(\tau_{\mathrm{SW}} = 0.05\) & \(O(nP \log n)\); \(P = 100\) \\
\botrule
\end{tabular}
\footnotetext[1]{Chi-square, Jensen--Shannon, Kullback--Leibler, PSI, KS, and univariate Wasserstein-1, applied per feature.}
\end{table}

\section{Problem Formulation and Two-Sample Testing Methodology}\label{sec3}

\subsection{Two-Sample Detection and Test Families}\label{subsec3-1}

Let \(\mathcal{D}_R = \{z_i\}_{i=1}^{n_R}\) and \(\mathcal{D}_C = \{z_j\}_{j=1}^{n_C}\) be a reference and a current sample drawn from distributions \(P_R\) and \(P_C\), where \(z = (X, Y)\) for multi-column tests and \(z = X_k\) for univariate marginal tests. Drift detection evaluates whether the two samples originate from identical distributions via the hypothesis:
\begin{equation}\label{eq:batch-null}
H_0 : P_R = P_C \qquad \text{vs.} \qquad H_1 : P_R \neq P_C.
\end{equation}
A batch protocol evaluates this via a test statistic \(s = T(\mathcal{D}_R, \mathcal{D}_C)\) and, where a null distribution is available, a \(p\)-value \(p\). The decision is \(\psi = \indicator[p < \alpha_0]\) for tests calibrated against a null distribution (permutation or asymptotic, with Bonferroni correction where applicable), and \(\psi = \indicator[s > \tau]\) for statistics compared against a fixed or null-calibrated threshold \(\tau\) (Table~\ref{tab:tests-overview}). Under \(H_0\) the pooled sample is exchangeable, which is the assumption underlying the permutation nulls of Section~\ref{subsec4-1}. We employ two structural test families to evaluate this hypothesis. \emph{Univariate marginal tests} independently compare the marginals \(P_R(X_k)\) and \(P_C(X_k)\) for each feature dimension \(k\), remaining strictly blind to changes in correlation structure. \emph{Multi-column tests} operate on the joint stacked vector, which in our operational framing incorporates both the features and the label as \((X, Y)\); the label is model-emitted in deployment, the generator class in the Harvard Dataverse streams, and \(Y_{\text{synth}}\) in the Trendyol benchmark. The multi-column tests span three mechanistic sub-families: (1) marginal per-feature aggregations (e.g., per-dimension KS with Bonferroni), (2) projection-based tests (sparse random projection KS, Sliced Wasserstein Distance), and (3) genuinely multivariate kernel embeddings (MMD-RBF, RFF-MMD). Table~\ref{tab:notation} (Appendix~\ref{secA1}) summarizes the notation used throughout the paper.

\subsection{Sliding-Window Changepoint Detection}\label{subsec3-2}

Because production drift is a continuous process rather than a single-threshold event, batch detection is insufficient for temporal localization. We extend the formulation into a sliding-window framework by partitioning the stream-ordered current dataset into \(N\) contiguous position buckets:
\begin{equation}\label{eq:bucket-partition}
    \mathcal{D}_C = \bigsqcup_{b=0}^{N-1} \mathcal{B}_b,
    \qquad
    \mathcal{B}_b = \{z_i : \lfloor b\,n_C/N \rfloor \leq i < \lfloor (b{+}1)\,n_C/N \rfloor\}.
\end{equation}
At each normalized position \(b/N \in [0, 1)\), the test yields a statistic \(s_b = T(\mathcal{D}_R, \mathcal{B}_b)\), forming a trajectory matrix \(\mathbf{S}\) across the stream. We implement two operational variants of this framework, corresponding to the sliding and fixed reference-window types distinguished by \citet{hinder2024survey}. The \emph{moving-reference} variant slides both windows synchronously to compare the immediate current window against the near past (applied to the Harvard Dataverse): at stream index \(t\) it compares the reference window \(\{z_i : t-W \leq i < t\}\) with the current window \(\{z_i : t \leq i < t+W\}\), advancing \(t\) by a stride \(\Delta\). Conversely, the \emph{fixed-reference} variant (applied to the Trendyol benchmark) compares each stream-ordered current bucket \(\mathcal{B}_b\) against a static, pre-drift reference set \(\mathcal{D}_R\). The latter isolates trajectory changes strictly to the current stream and substantially reduces distributed compute costs. 

\subsection{Ground-Truth Aggregation and Multiple Testing}\label{subsec3-3}

To evaluate trajectory quality beyond binary True Positive/False Positive rates, synthetic protocols allow us to aggregate ground-truth metrics per bucket. By computing the expected drift magnitude (\(\bar m_b\)) and mean drift intensity (\(\bar\alpha_b\)) for each \(\mathcal{B}_b\), we can evaluate the continuous correlation between a test's statistic trajectory and the true underlying drift severity, independent of threshold calibration. 

Calibration and multiple testing require careful handling in this regime. Permutation-based tests (MMD-RBF, RFF-MMD) natively utilize \(\alpha_0 = 0.05\). Bonferroni-aggregated tests apply a corrected threshold \(\alpha_0/d\) to maintain the family-wise error rate (FWER); however, statistical power degrades rapidly as dimensionality \(d\) increases. Furthermore, sliding-scan protocols inherently introduce cross-position dependence---either through overlapping current windows or a shared fixed reference---complicating the operational meaning of nominal \(\alpha_0\). Consequently, trajectory results in our continuous protocols are reported descriptively (hit rate, detection latency, stationary FPR) rather than relying on independence assumptions that do not hold in sequential scans.

\section{Algorithmic Framework and Distributed Big Data Execution}\label{sec4}

This section details the algorithmic implementation of the two-sample tests introduced in Section~\ref{subsec3-1} and outlines their execution architecture for production-scale data. To facilitate direct comparison across families, thresholding strategies, and computational complexities, the algorithmic configurations are synthesized in Table~\ref{tab:tests-execution}.

\subsection{Algorithmic Implementations and Calibration}\label{subsec4-1}

We implement tests across the three established sub-families: marginal aggregation, projection-based, and genuinely multivariate kernel embeddings. The theoretical scaling constraints of these algorithms fundamentally dictate their calibration strategies and deployment viability. For instance, the exact Maximum Mean Discrepancy with an RBF kernel (MMD-RBF) scales as \(O(n^2 d)\) due to kernel-matrix materialization, capping its practical application to a few thousand subsampled rows. To achieve linear time \(O(n \cdot D \cdot d)\) scaling suitable for the \({\sim}137.5\)M-row Trendyol stream, we utilize Random Fourier Features (RFF-MMD) \citep{rahimi2007random}, projecting inputs into a \(D\)-dimensional space (\(D = 500\) in the Failing Loudly runs and \(D = 200\) in the Trendyol scans; Section~\ref{subsubsec6-3-1}). 

Decision calibration closely follows the test constraints. Non-parametric permutation nulls are utilized for the kernel methods (MMD-RBF, RFF-MMD) to maintain strict size control under exchangeability assumptions. Conversely, Bonferroni-aggregated tests adjust the threshold by the effective dimensions (\(\alpha_0/d\) or \(\alpha_0/K\)), and Sliced Wasserstein Distance (SWD) relies on a fixed distance threshold (\(\tau_{\mathrm{SW}} = 0.05\)) without a permutation null, acting as a distance-based signal rather than a size-controlled test.

\begin{table}[!htbp]
\centering
\caption{Summary of Evaluated Two-Sample Tests and Execution Architecture}
\label{tab:tests-execution}
\footnotesize
\begin{tabular}{@{}>{\raggedright\arraybackslash}p{0.17\textwidth}>{\raggedright\arraybackslash}p{0.12\textwidth}>{\raggedright\arraybackslash}p{0.25\textwidth}>{\raggedright\arraybackslash}p{0.17\textwidth}>{\raggedright\arraybackslash}p{0.15\textwidth}@{}}
\toprule
\textbf{Test Name} & \textbf{Sub-Family} & \textbf{Core Statistic / Mechanism} & \textbf{Threshold / Calibration} & \textbf{Execution Paradigm} \\ \midrule
\textbf{Univariate (6 variants)} & Marginal & Marginal \(P_R(X_k)\) vs \(P_C(X_k)\) per feature & Fixed \(\alpha_0=0.05\) or lit. \(\tau\) & Hybrid (Driver aggregated) \\
\textbf{Per-dim KS} & Marginal Agg. & \(\max_k D_k\) across \(d\) raw dimensions & Bonferroni (\(\alpha_0/d\)) & Driver-side (Subsampled) \\
\textbf{Projected KS (SRP)} & Projection & \(\max\) KS across \(K=32\) sparse random projections & Bonferroni (\(\alpha_0/K\)) & Driver-side (Subsampled) \\
\textbf{Sliced Wasserstein Distance} & Projection & Avg. Wasserstein-1 over 100 random unit directions & Fixed distance (\(\tau = 0.05\)) & \textbf{Spark Distributed} \\
\textbf{MMD-RBF} & Kernel Embed. & \(\widehat{\mathrm{MMD}}^2\) via Gaussian RBF (\(O(n^2 d)\)) & Permutation (\(B=1000\)) & Driver-side (Subsampled) \\
\textbf{RFF-MMD} & Kernel Embed. & \(\|\bar{\mathbf{h}}(X_R) - \bar{\mathbf{h}}(X_C)\|_2^2\) mapping (\(D=500\)) & Permutation (\(B=200\)) & \textbf{Spark Distributed} \\ \bottomrule
\end{tabular}
\end{table}

\subsection{Spark Cluster Configuration and Execution Mapping}\label{subsec4-2}

All distributed computations are executed on Google Cloud Dataproc Serverless (PySpark 3.5), utilizing up to 16 executors (16 vCPU, 96\,000~MiB Java virtual machine heap per executor). Because the statistical tests exhibit vastly different memory properties, execution maps onto two distinct paths, as indicated in Table~\ref{tab:tests-execution}:

\begin{itemize}
    \item \textbf{Spark-Distributed Execution:} Tests inherently scalable to \(O(N)\) (RFF-MMD and SWD) compute their core statistics via distributed transformations across the full reference and current buckets. RFF-MMD distributes the \(D\)-dimensional Fourier feature computation and mean embedding aggregation, while SWD distributes the random unit projections and quantile approximations. Execution time is $\sim$2.7--5.6~s per position.
    \item \textbf{Driver-Side Execution:} Tests bottlenecked by matrix materialization or lacking distributed native implementations (MMD-RBF, Projected KS, Per-dim KS) require the bucket to be collected to the driver via a limit query (driver sample size \(n_{\mathrm{drv}} = 20{,}000\); further subsampled to 3,000 for MMD-RBF). Null shuffles and linear algebra operations are computed locally in-memory on the driver.
\end{itemize}

This architectural split separates tests whose cost scales with the full bucket from tests that operate on a driver-side subsample; Section~\ref{subsubsec6-3-8} quantifies the resulting runtime gap.

\section{Experimental Design and Benchmark Protocols}\label{sec5}

\subsection{Dataset Inventory and Preparation}\label{subsec5-1}

We evaluate the detection methods across three structural layers, each providing a distinct type of ground truth and validation claim (Table~\ref{tab:dataset-inventory}). The \textbf{Harvard Dataverse} streams \citep{lopez2020streams} provide controlled conditional shifts (\(P(Y \mid X)\) changes, \(P(X)\) is stationary). The \textbf{Failing Loudly} image-shift benchmark \citep{rabanser2019failing} provides static, heavily-shifted covariate evaluation environments. Finally, the \textbf{Trendyol collection-ranking feature table} provides massive scale (\({\sim}137.5\) million rows) and authentic user-behavior variance. The table is hosted in Google BigQuery, Google Cloud's serverless, columnar data warehouse for SQL analytics over petabyte-scale datasets. Trendyol's data originates from live A/B tests on a regional-leader e-commerce recommendation system; following the pre-check (Appendix~\ref{secA2}), the benchmark operates on an 11-dimensional vector of metadata scalars (identifier columns retained by design) and live similarity metrics.

\begin{table}[!htbp]

\caption{Dataset inventory and feature dimensionality. Feature dimensionality is reported after one-hot encoding of categorical covariates.}\label{tab:dataset-inventory}
\centering
\footnotesize
\begin{tabular}{@{}>{\raggedright\arraybackslash}p{0.18\textwidth}>{\raggedright\arraybackslash}p{0.27\textwidth}rr>{\raggedright\arraybackslash}p{0.24\textwidth}@{}}
\toprule
\textbf{Layer} & \textbf{Source / Family} & \textbf{Configs} & \textbf{Dim} & \textbf{Ground Truth} \\ \midrule
\textbf{Harvard Dataverse} & SEA, SINE, STAGGER, MIXED, Random Tree  & 20 & 2--9 & 3 temporal changepoints \\ \addlinespace
\textbf{Failing Loudly}    & MNIST, CIFAR-10 (10 shift types)   & 60 & 784 / 3072 & Image perturbation labels \\ \addlinespace
\textbf{Trendyol benchmark}    & Collection-ranking feature table & 8 & 11 & Injected row-level drift (Eqs.~\ref{eq:eligible}--\ref{eq:realized-flip}) \\
\botrule
\end{tabular}
\end{table}

\subsection{Evaluation Protocols}\label{subsec5-2}

\begin{figure}[!htbp]
\centering
\includegraphics[width=0.8\textwidth]{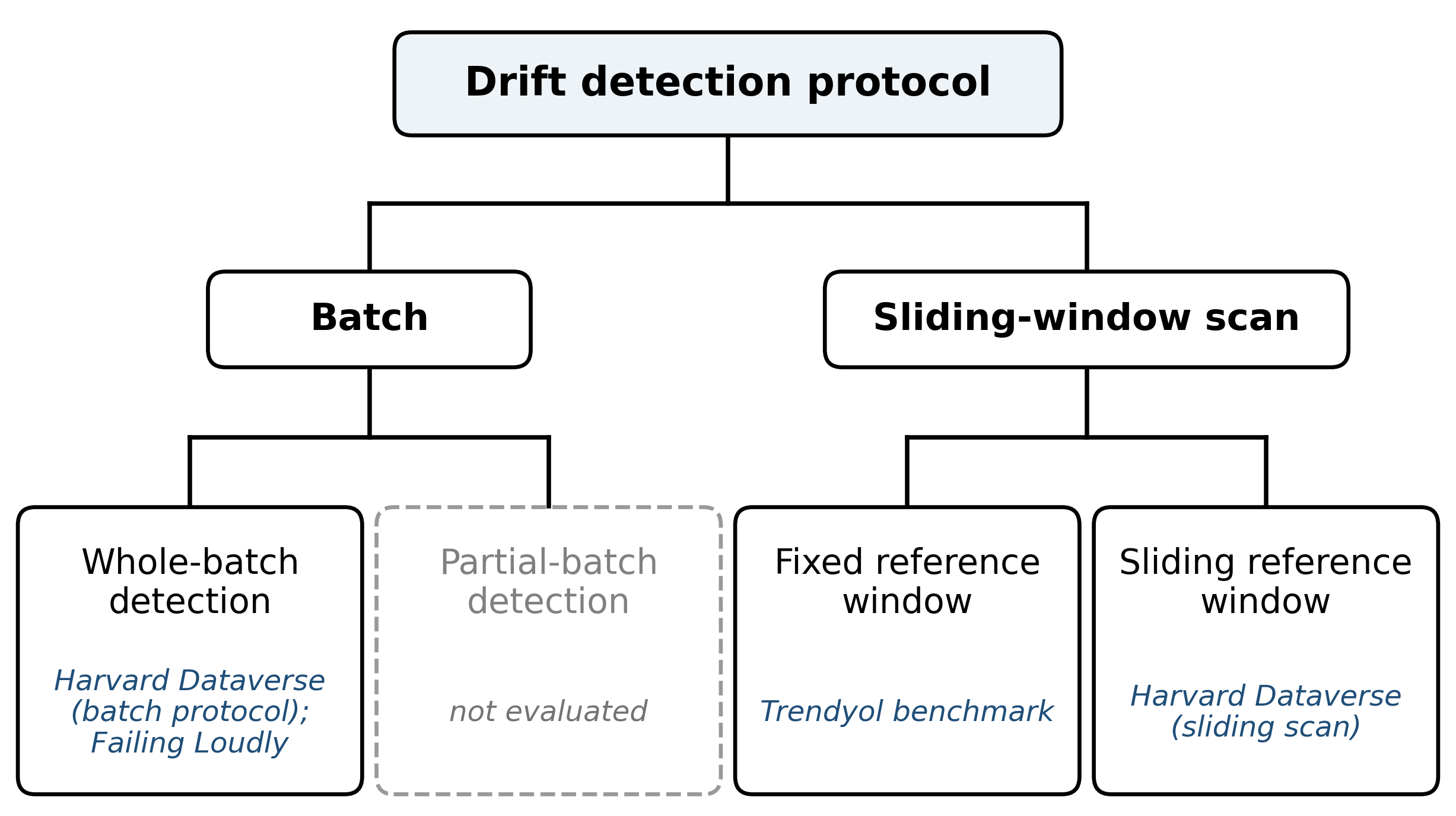}
\caption{Evaluation protocols of this study within the taxonomy of unsupervised drift detectors of \citet{gemaque2020overview} (adapted from their Fig.~2, CC BY 4.0). Batch detectors test an accumulated window as a whole; sliding-window scans advance the detection window along the stream and differ in whether the reference window stays fixed or slides with it. The survey's online detectors check every arriving instance, whereas our scans evaluate at a stride \(\Delta\) (Harvard Dataverse) or per position bucket (Trendyol benchmark). Partial-batch detection, which tests only a selected subset of each batch, is not evaluated here.}\label{fig:detection-taxonomy}
\end{figure}

To systematically assess the test families, we apply three distinct protocols, summarized in Table~\ref{tab:protocol-summary}. The \emph{Batch} protocol identifies the binary presence of drift, the \emph{Sliding-Scan} protocol evaluates temporal localization and detection latency, and the \emph{Failing Loudly Replication} validates our implementations against published literature baselines by mean absolute error (MAE). 
Figure~\ref{fig:detection-taxonomy} positions these protocols within the taxonomy of unsupervised drift detectors proposed by \citet{gemaque2020overview}.

\begin{table}[!htbp]
\caption{Summary of experimental protocols and evaluation metrics.}\label{tab:protocol-summary}
\centering
\footnotesize
\begin{tabular}{@{}>{\raggedright\arraybackslash}p{0.17\textwidth}>{\raggedright\arraybackslash}p{0.29\textwidth}>{\raggedright\arraybackslash}p{0.22\textwidth}>{\raggedright\arraybackslash}p{0.20\textwidth}@{}}
\toprule
\textbf{Protocol} & \textbf{Data Structure \& Windows} & \textbf{Evaluated Tests} & \textbf{Primary Metrics} \\ \midrule
\textbf{Batch} (Harvard) & Ref: 1st half. Cur: 2nd half. & All 11 tests (Marginal + Joint) & Binary Recall (TP / TP+FN) \\ \addlinespace
\textbf{Sliding-Scan} (Harvard) & \(W = 5000\), stride \(\Delta = 1000\). Moving reference \([t-W, t-1]\). & All 11 tests & Hit rate (\(\rho=200\)), Latency, FPR \\ \addlinespace
\textbf{Sliding-Scan} (Trendyol) & \(N = 50\) position buckets. Fixed reference table (first 100,000 rows; 10,000 for driver-side tests). 8 scenarios. & MMD-RBF, Per-dim KS, SRP+KS, RFF-MMD, SWD & Pearson \(r\) vs.\ \(\bar m_b\), TPR, FPR, Runtime \\ \addlinespace
\textbf{Failing Loudly} & Static sizes \(N \in \{10 \dots 10,000\}\). 5 random splits, 3 fractions. & MMD-RBF, Per-dim KS, SRP+KS, RFF-MMD, SWD & \(\mathrm{Acc}(N)\), MAE vs literature \\
\bottomrule
\end{tabular}
\end{table}

\subsection{Synthetic Concept Drift Injection on Production Data}\label{subsec5-4}

Because the Trendyol collection-ranking feature table lacks stationary ground truth, we construct a controlled evaluation environment via a synthetic concept drift injection. We define two deterministic concepts based on live similarity metrics: Concept A (visual similarity, \(P(Y_A=1) \approx 0.22\)) and Concept B (textual similarity, \(P(Y_B=1) \approx 0.45\)). The injection overlays a synthetic label \(y'\) without altering the original row-wise features (\(X_{\text{after}} = X_{\text{before}}\)), ensuring any detected signal is strictly a change in \(P(Y \mid X)\).

\begin{table}[!htbp]
\caption{Trendyol Benchmark Scenario Configurations. The strong regime measures clean detection capabilities, while the weak regime establishes an empirical low-signal failure point.}\label{tab:scenario-config}
\centering
\small
\begin{tabular*}{\textwidth}{@{\extracolsep\fill}l cc c c @{}}
\toprule
\textbf{Drift Type (Regime)} & \textbf{Severity (\(\sigma\))} & \textbf{Scope (\(\phi\))} & \textbf{Mean \(\bar\alpha_s\)} & \textbf{\(\text{FlipRate}_{\text{scenario}}^{\text{expected}}\)} \\ \midrule
\textbf{sudden\_strong}      & 0.75 & 0.60 & 0.400 & 10.3\% \\
\textbf{sudden\_weak}        & 0.25 & 0.10 & 0.400 &  0.57\% \\
\textbf{gradual\_strong}     & 0.75 & 0.60 & 0.325 &  8.4\% \\
\textbf{gradual\_weak}       & 0.25 & 0.10 & 0.325 &  0.47\% \\
\textbf{incremental\_strong} & 0.75 & 0.60 & 0.325 &  8.4\% \\
\textbf{incremental\_weak}   & 0.25 & 0.10 & 0.325 &  0.47\% \\
\textbf{recurring\_strong}   & 0.75 & 0.60 & 0.200 &  5.2\% \\
\textbf{recurring\_weak}     & 0.25 & 0.10 & 0.200 &  0.29\% \\
\botrule
\end{tabular*}
\end{table}

The transition is controlled by three parameters: the temporal blend weight \(\alpha \in [0, 1]\), the scope fraction \(\phi\) (rows exposed), and the severity \(\sigma\) (mixing strength). The active target probability per row is:
\begin{equation}\label{eq:active-target-prob}
    p_{\text{active}}(\text{row}) = p_A(\text{row}) + \alpha \cdot \left[p_B^{\text{effective}}(\text{row}) - p_A(\text{row})\right]
\end{equation}
where \(p_B^{\text{effective}}(\text{row}) = p_A(\text{row}) + \sigma \cdot [p_B(\text{row}) - p_A(\text{row})]\) if the row is in scope, and \(p_A(\text{row})\) otherwise. The injected label is drawn via \(y'(\text{row}) = \indicator[U(0,1) < p_{\text{active}}(\text{row})]\). Throughout, \(y_A(\text{row}) = Y_A(\text{row})\) denotes the pre-drift label and \(y'(\text{row})\) the injected label; \(Y_{\text{synth}}\) denotes the corresponding label variable in the joint vector \((X, Y_{\text{synth}})\). 

We formulate eight production scenarios (four drift types \(\times\) two regimes) to stress-test the algorithms (Table~\ref{tab:scenario-config}). The expected realized-flip fraction for a scenario (\(\text{FlipRate}\)) and the fully-active-bucket expected magnitude (\(m_{\text{bucket}}^{\text{active}}\)) quantify the drift intensity:
\begin{align}
\label{eq:scenario-flip-rate}
    \text{FlipRate}_{\text{scenario}}^{\text{expected}} &\approx \bar\alpha_s \cdot \sigma \cdot \phi \cdot \delta \\
\label{eq:bucket-magnitude}
    m_{\text{bucket}}^{\text{active}} &\approx \sigma \cdot \phi \cdot \delta 
\end{align}
where \(\bar\alpha_s\) is the mean temporal coefficient and \(\delta \approx 0.5724\) is the Concept A/B disagreement rate.

To correlate a test's statistic trajectory with the true drift severity independently of threshold calibration, the injection framework also emits three row-level ground-truth variables: the eligibility indicator \(e\), the expected drift magnitude \(m\), and the realized-flip indicator \(f\) (the corresponding stored columns are listed in Table~\ref{tab:feature-glossary}):
\begin{align}
    \label{eq:eligible}
    e(\text{row}) &= \indicator[\text{in\_scope} \wedge Y_A(\text{row}) \neq Y_B(\text{row})], \\
    \label{eq:expected-magnitude}
    m(\text{row}) &= \alpha(\text{row}) \cdot \sigma \cdot e(\text{row}), \\
    \label{eq:realized-flip}
    f(\text{row}) &= \indicator[y'(\text{row}) \neq y_A(\text{row})].
\end{align}
The per-bucket means \(\bar m_b\) and \(\bar f_b\) of \(m\) and \(f\) over the rows of \(\mathcal{B}_b\) serve as the primary and robustness correlation targets, respectively; \(\bar m_b\) is preferred because it precisely encodes the probability-distribution displacement without the variance introduced by per-row Bernoulli sampling noise.

\section{Results}\label{sec6}

\subsection{Harvard Dataverse Concept Drift Streams}\label{subsec6-1}

This section evaluates the batch protocol (defined in Section~\ref{subsec5-2}) across the 20 Harvard Dataverse streams to determine if stream-level shifts are detectable; Section~\ref{subsubsec6-1-2} reports the corresponding sliding-scan results. Because these streams are engineered with a stationary marginal \(P(X)\) while only the conditional \(P(Y \mid X)\) shifts, they provide a rigorous diagnostic environment for distinguishing joint-distribution sensitivity from purely marginal sensitivity.

Because recall is either \(0\%\) or \(100\%\) for every test--family pair, Table~\ref{tab:harvard-batch-recall} summarizes the batch outcomes by test category. The results isolate three critical behavioral bifurcations across the test families:

\begin{enumerate}
    \item \textbf{Robustness of Genuine Multivariate Tests:} Projection-based and kernel embedding tests (RFF-MMD, SWD, MMD-RBF, Projected KS) achieve \(100\%\) recall on all five stream families. This empirically validates their theoretical capacity to detect shifts in the joint distribution \((X, Y)\) even when individual feature marginals remain perfectly stationary.
    
    \item \textbf{Structural Failure of Per-dim KS:} Despite operating on the stacked vector (features \(\cup\) class), Bonferroni-aggregated Per-dim KS completely fails (\(0\%\) recall) on the SINE and MIXED families. This highlights a severe operational vulnerability: the test's inherent blindness to pure dependence changes, entangled with Bonferroni conservativeness, eliminates its detection power in specific conditional-shift regimes.
    
    \item \textbf{Expected Failure of Univariate Marginals:} Categorical marginal tests (chi-square, JS-div, KL-div, PSI) exhibit \(0\%\) recall on four of the five families, precisely as analytically expected under stationary \(P(X)\). The sole exception is the Random Tree (RT) family, where an artifact in the stream generator rescales the one-hot encoded class column at changepoints, inadvertently creating a detectable marginal signal.
\end{enumerate}

\begin{table}[!htbp]
\centering
\caption{Aggregated Batch Recall on Harvard Dataverse Streams. The bifurcation in recall demonstrates the structural failure of marginal tests under stationary \(P(X)\) environments, contrasted against the robustness of joint multivariate embeddings.}
\label{tab:harvard-batch-recall}
\footnotesize
\begin{tabular}{@{}>{\raggedright\arraybackslash}p{0.30\textwidth}ccccc@{}}
\toprule
\textbf{Test Category} & \textbf{SEA} & \textbf{SINE} & \textbf{STAGGER} & \textbf{MIXED} & \textbf{Random Tree (RT)} \\ \midrule
\textbf{Multivariate (Kernel \& Proj.)} & 100\% & 100\% & 100\% & 100\% & 100\% \\
\textbf{Per-dim KS (Bonferroni)} & 100\% & 0\% & 100\% & 0\% & 100\% \\
\textbf{Univariate Marginal} & 0\% & 0\% & 0\% & 0\% & 100\%\footnotemark[1] \\ \bottomrule
\end{tabular}
\footnotetext[1]{Recall artifact resulting from a generator-induced class-marginal rescaling at changepoints.}
\end{table}

\begin{figure*}[!htbp]
\centering
\includegraphics[width=\textwidth]{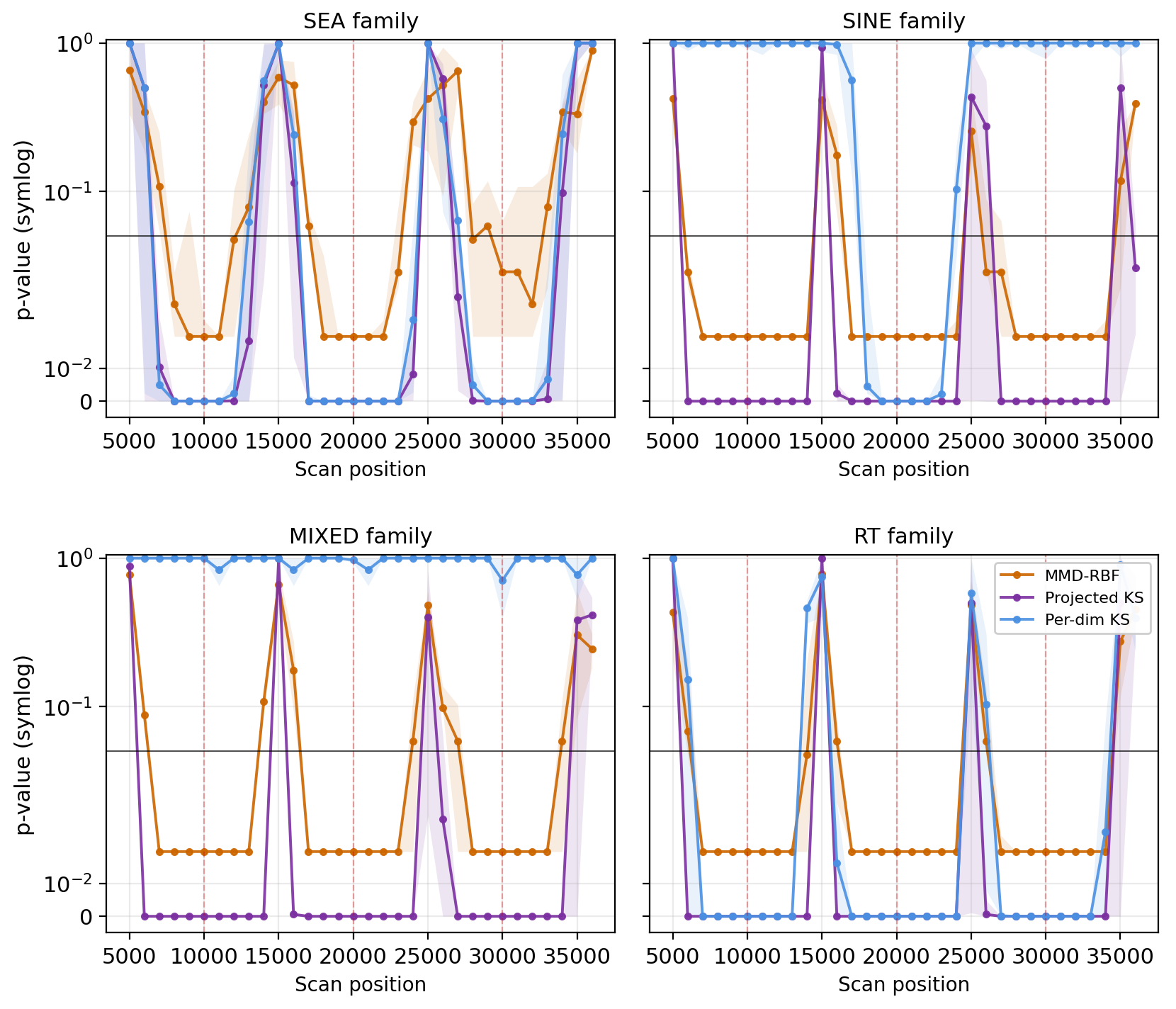}
\caption{Sliding-scan \(p\)-value trajectories on Harvard Dataverse streams. Each panel aggregates streams within one family and plots the median \(p\)-value at each scan position for MMD-RBF, projected KS, and per-dim KS. Shaded bands are interquartile ranges over streams. Red dashed lines mark ground-truth changepoints; the horizontal line is \(\alpha_0 = 0.05\) on a symmetric-log axis. STAGGER is omitted because its whole-vector detection uses distance-based statistics without \(p\)-values after one-hot encoding of the categorical features.\label{fig:lopez-sliding-trajectories}}
\end{figure*}

\subsubsection{Sliding-scan results}\label{subsubsec6-1-2}

Figure~\ref{fig:lopez-sliding-trajectories} shows the sliding-scan trajectories (\(W = 5000\), \(\Delta = 1000\)) of the three \(p\)-value-producing tests on four stream families; STAGGER is omitted because the distance-based statistics used on its one-hot encoded features yield no \(p\)-values. MMD-RBF gives the cleanest structure: in the SEA, MIXED, and RT families its \(p\)-value falls below \(\alpha_0 = 0.05\) around every changepoint and stays high in stationary regions, capturing all changepoints within the \(\rho = 200\) tolerance at a typical latency of 1--2 sample steps, small relative to the \(W = 5000\) window. Per-dim KS reproduces its batch failure: on SINE its \(p\)-value stays near 1 throughout the scan, and it captures no changepoint in SINE or MIXED. Projected KS lies in between. Projecting to \(d = 32\) dimensions eases the conservativeness of the Bonferroni correction, but information mixing weakens the signal at some changepoints, and it succeeds only partially, in two families.

\FloatBarrier 
\subsection{Failing Loudly Reproduction and Image-Shift Results}\label{subsec6-2}

This section reports results of the Failing Loudly reproduction protocol defined in Section~\ref{subsec5-2}. Three tests (per-dim KS, projected KS with SRP, MMD-RBF) are validated against Failing Loudly Table~1(a) rows and characterized across the ten shift types, grouped into five families, at perturbation fractions \(\{10\%, 50\%, 100\%\}\); the two extension methods (RFF-MMD and SWD) are characterized under the same protocol in Appendix~\ref{secA5}.

\subsubsection{Reproduction accuracy}\label{subsubsec6-2-1}

The reproduced MMD-RBF detection accuracy at \(N = 1000\) deviates from the paper value by 0.007, and the reproduced per-dim KS deviation at \(N = 10\,000\) is 0.007. Mean absolute error (MAE) across the three methods over sample sizes lies in the range \([0.030, 0.053]\); this demonstrates that Rabanser et al.'s original measurements can be numerically reproduced. The largest single-cell deviations are for SRP+KS at \(N = 10\) (\(-0.140\)) and at \(N = 1000\) (\(-0.137\)); the former is consistent with the discreteness of the KS null distribution at \(N = 10\) under Bonferroni over 32 projected dimensions, and the latter has been examined in supplementary analysis but a fully quantitative explanation remains open.

\subsubsection{Shift-type by dataset breakdown}\label{subsubsec6-2-2}

\begin{figure*}[!htbp]
\centering
\includegraphics[width=\textwidth]{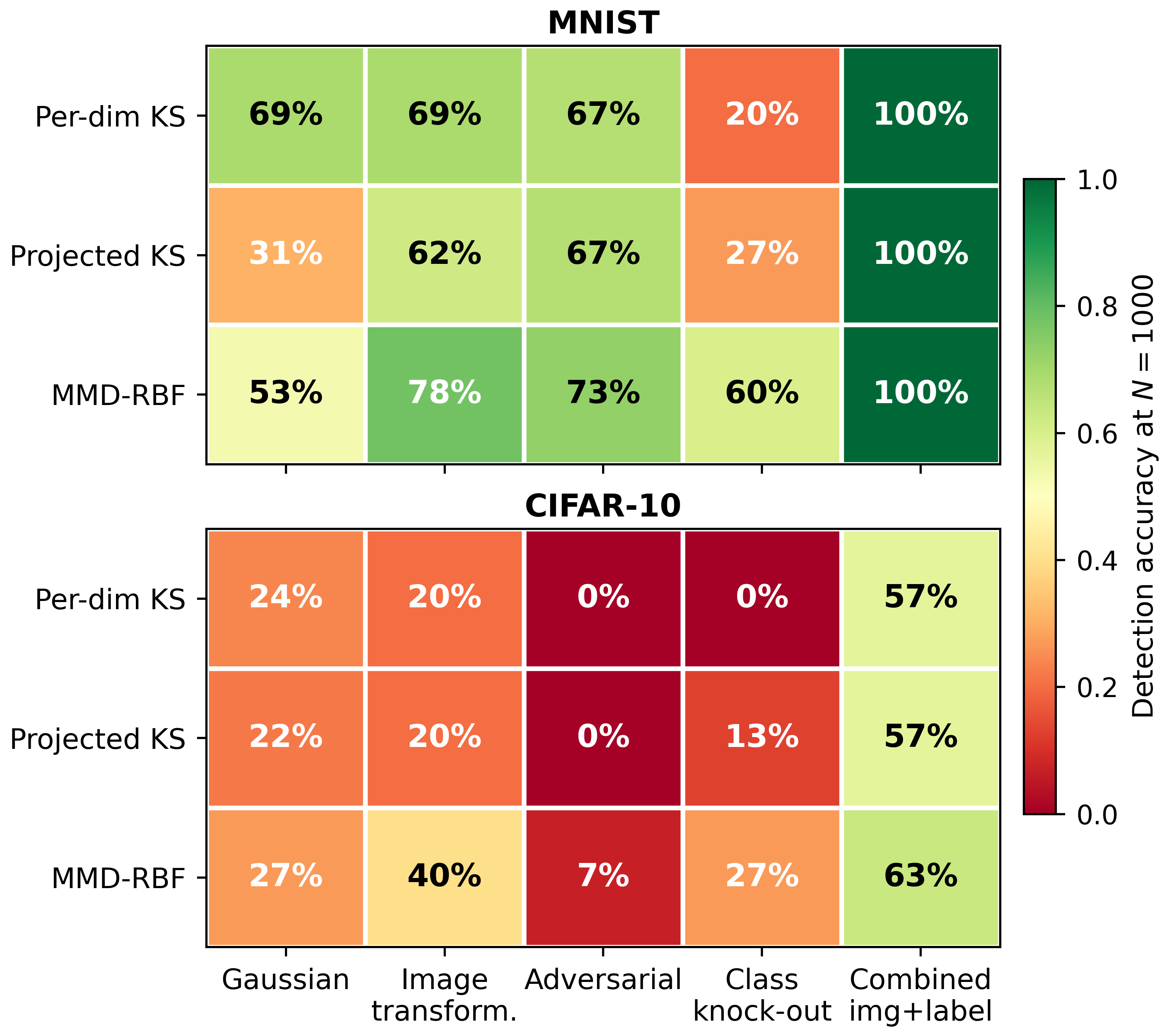}
\caption{Detection accuracy by test and shift family at \(N = 1000\) on the Failing Loudly benchmark, disaggregated by dataset. Rows are the three reproduced baselines (per-dimension KS on raw pixels, projected KS with SRP, MMD-RBF). Columns are the five shift families of Failing Loudly. Each cell aggregates all comparisons of one family over the three perturbation fractions and five splits: 45 for the Gaussian and image families (three intensities each), 30 for the combined family (two variants), and 15 for the adversarial and class-knockout families.\label{fig:fl-heatmap}}
\end{figure*}

Figure~\ref{fig:fl-heatmap} reports detection accuracy at \(N = 1000\) by shift type and dataset, revealing three patterns. Combined image and label shifts are the easiest: they alter both the covariate and the label distribution, and all three tests reach 100\% on MNIST and about 60\% on CIFAR-10. Adversarial shifts are the hardest: on CIFAR-10 both KS variants reach 0\% and MMD-RBF only 7\%, because adversarial perturbations target the decision boundary while leaving raw pixel marginals nearly unchanged. Class knock-out shifts fall in between. They change \(P(Y)\) without directly modifying \(P(X)\), yet removing classes alters the observed class-conditional covariate distribution, which MMD-RBF partly picks up (60\% on MNIST, 27\% on CIFAR-10) while the KS variants reach at most 27\% and 13\%, respectively.

\subsubsection{\texorpdfstring{\(p\)}{p}-value curves for Gaussian noise shifts}\label{subsubsec6-2-3}

\begin{figure*}[!htbp]
\centering
\includegraphics[width=\textwidth]{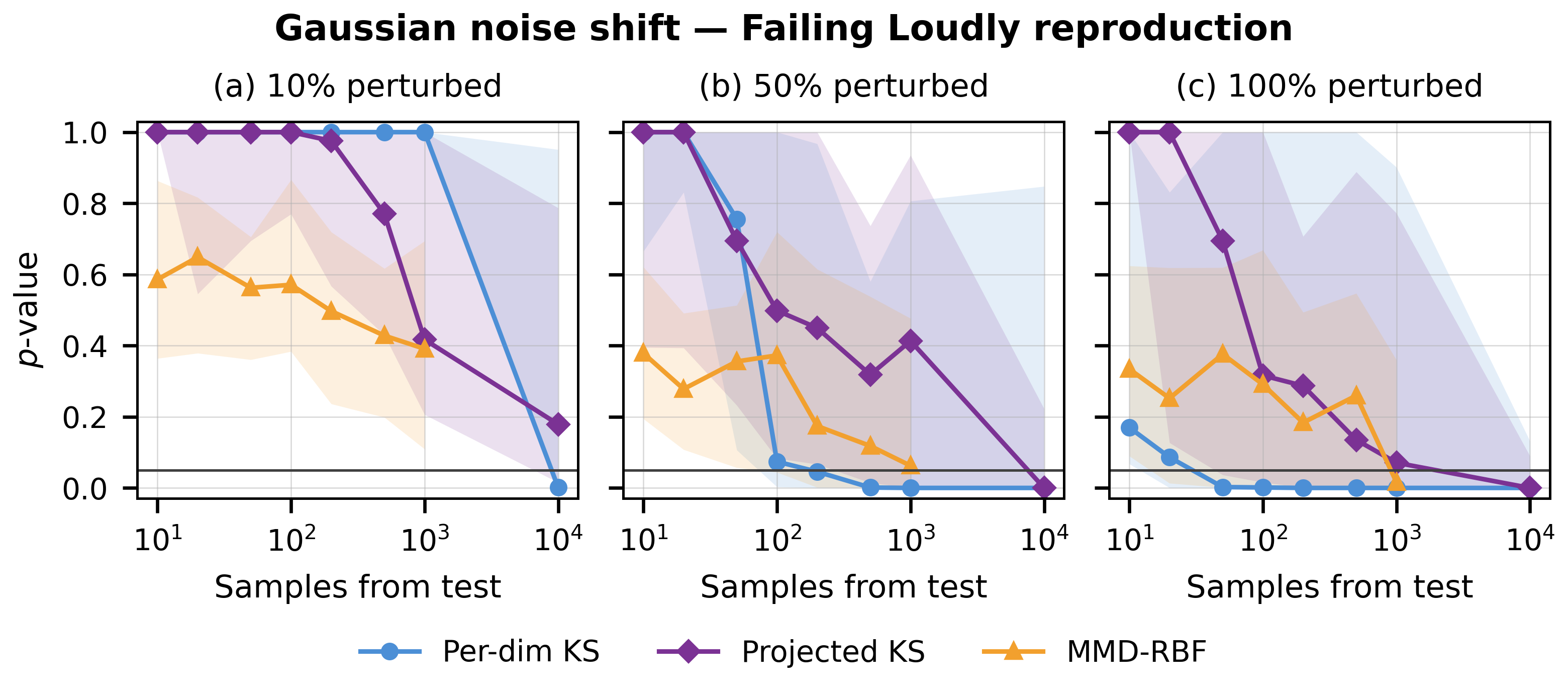}
\caption{Reproduced \(p\)-value curves under the Gaussian-noise shift class of \citet{rabanser2019failing} on MNIST and CIFAR-10. Panels correspond to affected-fraction \(\{0.10, 0.50, 1.00\}\). Lines are median \(p\)-values across 30 shift instances per sample size (3 intensities \(\times\) 5 splits \(\times\) 2 datasets). Shaded bands are the interquartile range. The horizontal reference line is \(\alpha_0 = 0.05\). This figure is the reproduction analog of Failing Loudly Figure~3.\label{fig:fl-gaussian}}
\end{figure*}

Figure~\ref{fig:fl-gaussian} shows the three methods' \(p\)-values vs.\ sample size for the Gaussian noise shift class at affected fractions \(\{0.10, 0.50, 1.00\}\) (the reproduction analog of Failing Loudly Figure~3; principal component analysis and extension methods omitted per Section~\ref{subsec5-2}); lines are medians pooled over both datasets, shift intensities, and splits. Detection depends strongly on the affected fraction. Per-dim KS falls below \(\alpha_0 = 0.05\) from \(N = 50\) at 100\% perturbation and from \(N = 200\) at 50\%, but at 10\% only at \(N = 10\,000\). MMD-RBF, evaluated up to \(N = 1000\), crosses \(\alpha_0\) only at 100\% perturbation and \(N = 1000\) (median \(p = 0.016\)); at 10\% its median stays near 0.4. Projected KS is the least sensitive: even at 100\% its median remains above \(\alpha_0\) up to \(N = 1000\) (\(p = 0.071\)) and falls below it only at \(N = 10\,000\), consistent with signal loss in the \(d = 32\) projected space.

\subsubsection{Additional shift types: image, adversarial, class-knockout, and combined}\label{subsubsec6-2-4}

\begin{figure*}[!htbp]
\centering
\includegraphics[width=\textwidth]{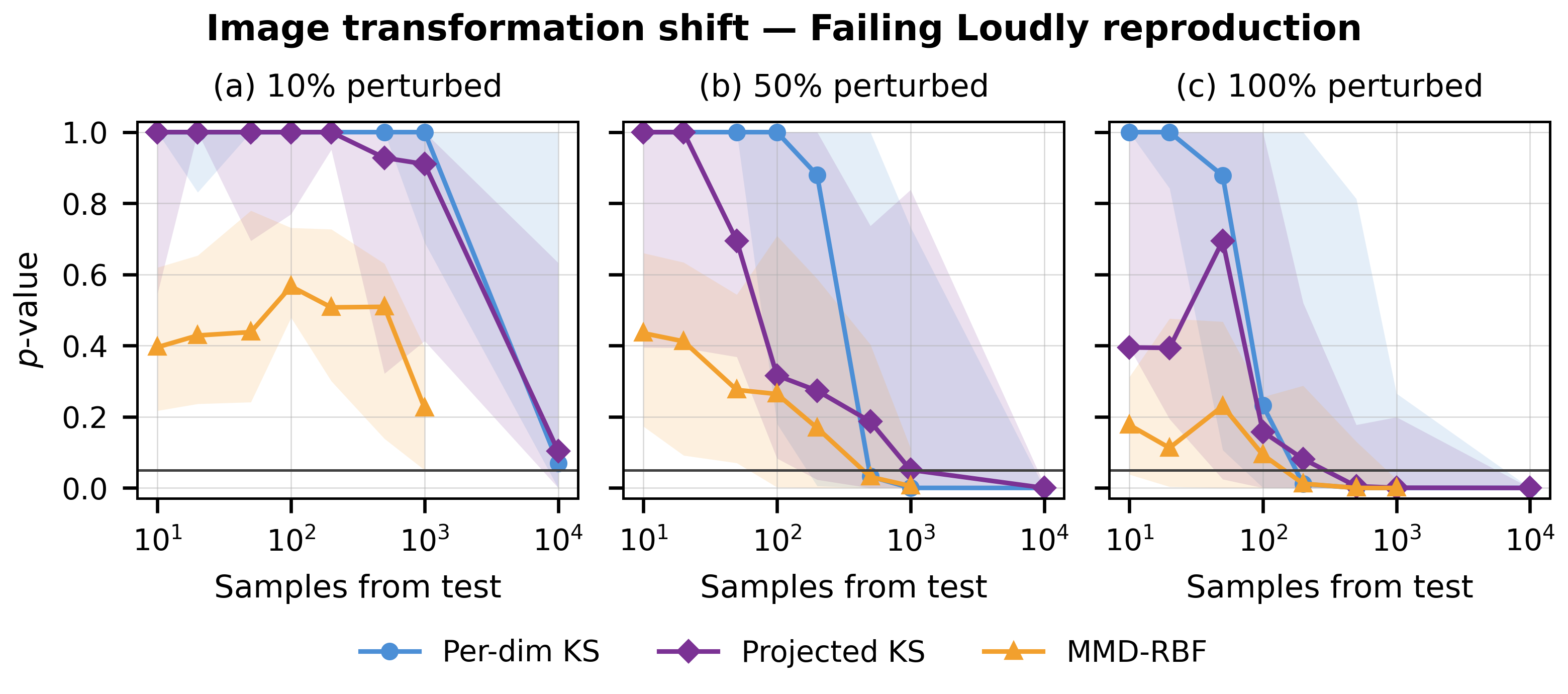}
\caption{Failing Loudly reproduction: image-transformation shifts (rotation, translation, zoom at three intensities). Panels correspond to \(\{10\%, 50\%, 100\%\}\) perturbed test data.\label{fig:fl-image}}
\end{figure*}

\begin{figure*}[!htbp]
\centering
\includegraphics[width=\textwidth]{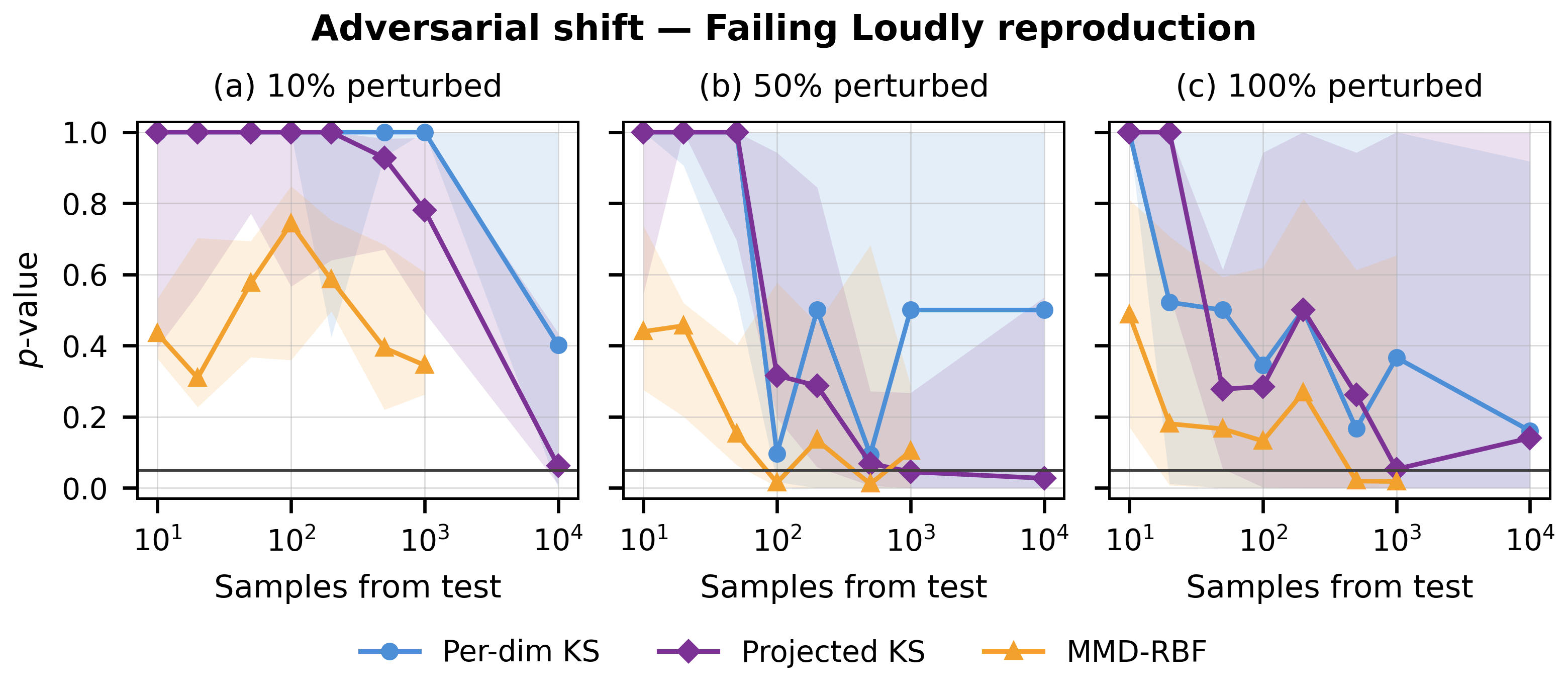}
\caption{Failing Loudly reproduction: adversarial shifts generated with the fast gradient sign method. Panels correspond to \(\{10\%, 50\%, 100\%\}\) perturbed test data. Adversarial shifts are the hardest single-class shift family at low perturbation fractions.\label{fig:fl-adversarial}}
\end{figure*}

\begin{figure*}[!htbp]
\centering
\includegraphics[width=\textwidth]{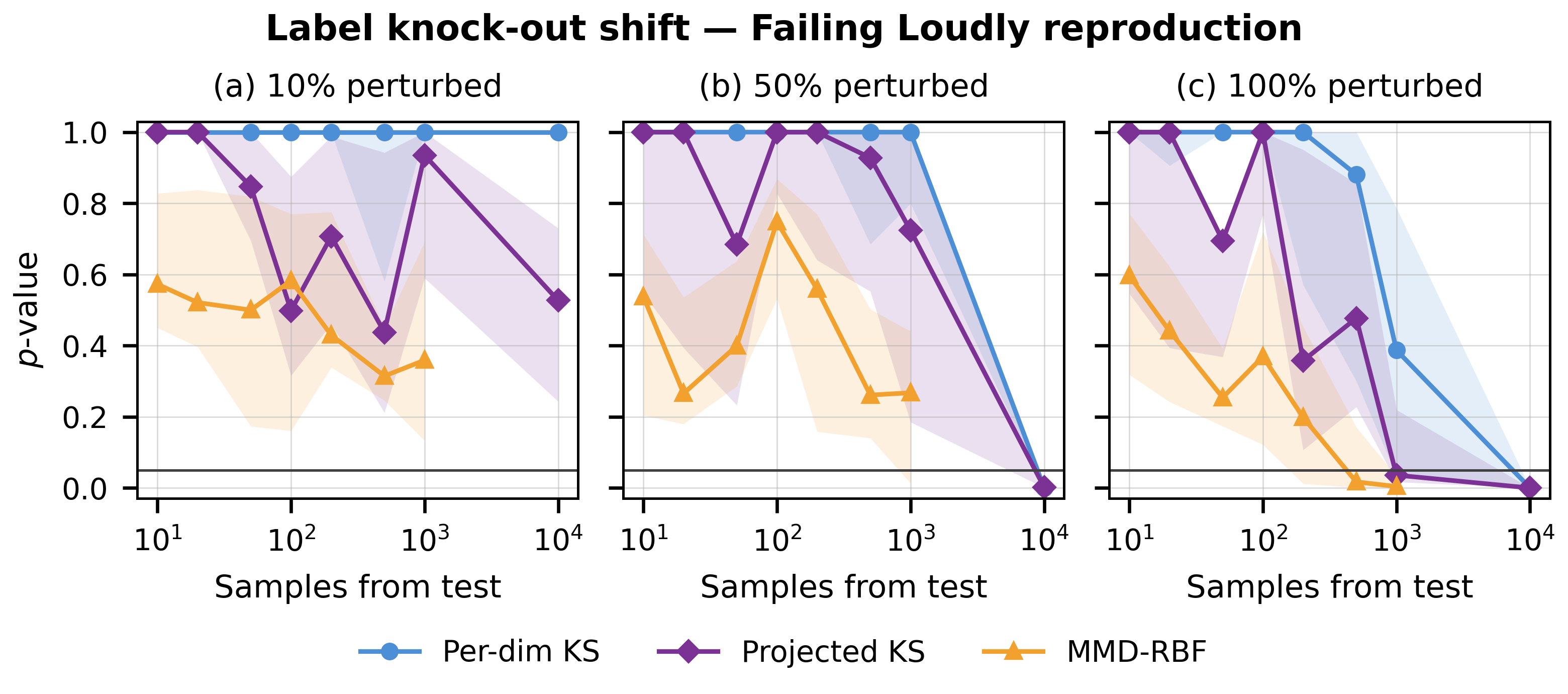}
\caption{Failing Loudly reproduction: class knock-out (label) shifts. Panels correspond to \(\{10\%, 50\%, 100\%\}\) perturbed test data. Because knock-out shifts change \(P(Y)\) rather than \(P(X)\), MMD-based tests detect the shift through its indirect effect on the observed covariate distribution.\label{fig:fl-label}}
\end{figure*}

\begin{figure*}[!htbp]
\centering
\includegraphics[width=\textwidth]{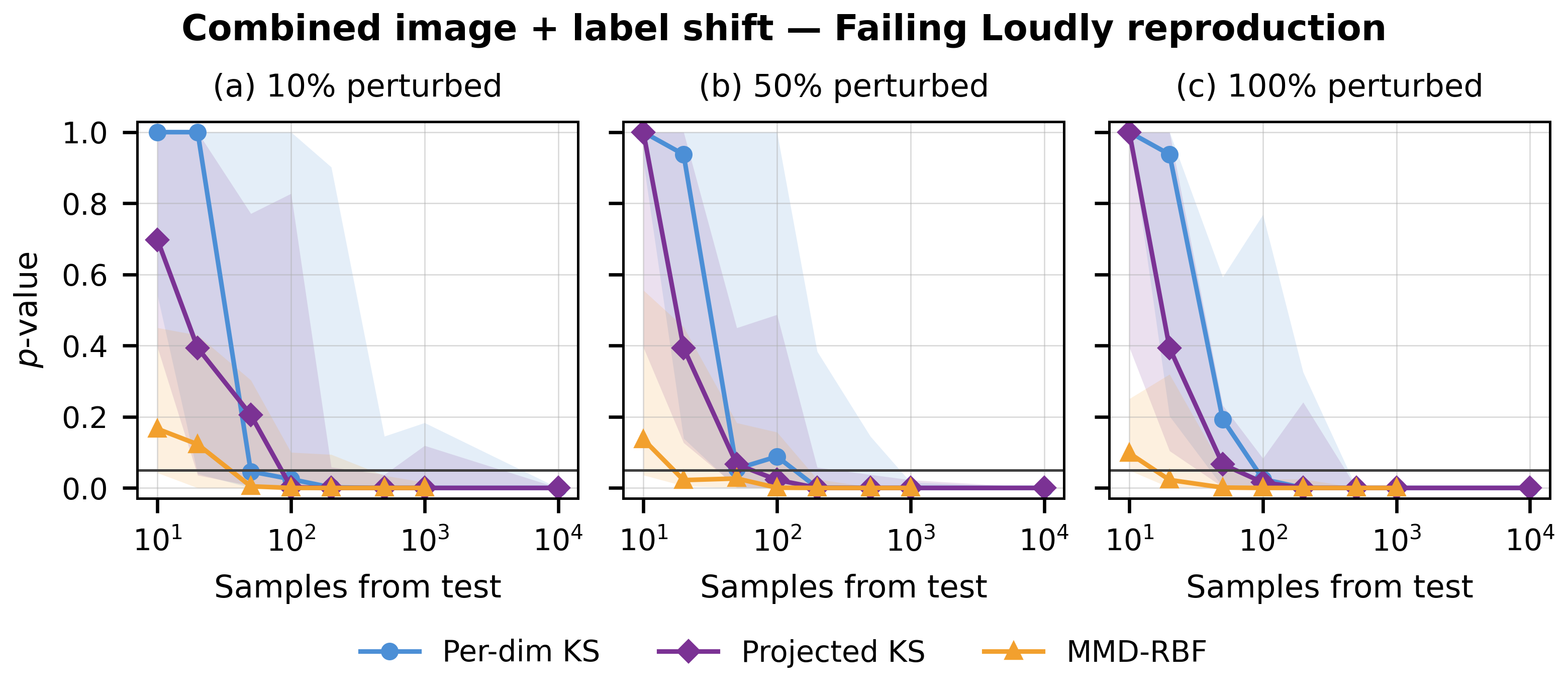}
\caption{Failing Loudly reproduction: combined image transformation + class knock-out shifts. Panels correspond to \(\{10\%, 50\%, 100\%\}\) perturbed test data. Combined shifts show the sharpest median-\(p\) descent because covariate and label distributions shift jointly.\label{fig:fl-combined}}
\end{figure*}

Beyond the Gaussian noise shifts of Section~\ref{subsubsec6-2-3}, the Failing Loudly protocol evaluates four additional shift families (Figures~\ref{fig:fl-image}--\ref{fig:fl-combined}), aggregated in the same way. \textbf{Image transformation shifts} (rotation, translation, zoom) behave like Gaussian noise: per-dim KS and MMD-RBF both fall below \(\alpha_0\) from \(N = 200\) at 100\% and from \(N = 500\) at 50\% perturbation, while at 10\% no test crosses \(\alpha_0\) at any evaluated sample size. \textbf{Adversarial shifts} are the hardest and show no consistent descent: at 10\% no test crosses \(\alpha_0\), and at 50\% and 100\% the median \(p\)-values fluctuate with \(N\) instead of decreasing monotonically. MMD-RBF reaches the lowest values (median \(p \approx 0.02\) at \(N = 500\)--\(1000\) under 100\% perturbation), whereas per-dim KS stays above \(\alpha_0\) at every sample size and fraction. \textbf{Class-knockout (label) shifts} keep the per-dim KS \(p\)-value well above \(\alpha_0\) up to \(N = 1000\) at every fraction (median \(p \geq 0.39\)); it falls below \(\alpha_0\) only at \(N = 10\,000\) for 50\% and 100\% perturbation. MMD-RBF picks up the indirect covariate effect only at 100\% perturbation, crossing \(\alpha_0\) from \(N = 500\) (median \(p = 0.017\)); at 10\% and 50\% its median remains above 0.25. \textbf{Combined image+label shifts} produce the sharpest descent because covariate and label distributions shift jointly: even at 10\% perturbation, per-dim KS and MMD-RBF fall below \(\alpha_0\) from \(N = 50\) and projected KS from \(N = 100\).

\FloatBarrier 
\subsection{Real-World Production Data with Synthetic Concept Drift Injection}\label{subsec6-3}

\subsubsection{Setup and threshold calibration}\label{subsubsec6-3-1}\label{subsubsec6-3-2}

The eight scenarios apply the fixed-reference sliding scan (Table~\ref{tab:protocol-summary}) to the 11-dimensional benchmark vector of Section~\ref{subsec5-1}. With \(s_0 = 0.6\) and 50 buckets of approximately equal size, bucket 29 straddles the drift onset and is the first bucket containing rows with \(\alpha > 0\) (\(\bar\alpha_{29} \approx 0.09\) under the sudden schedule). For sudden, gradual, and incremental drift, buckets 0--28 therefore form the injection-null region (\(\bar m_b = 0\)) and buckets 29--49 the injection-active region, so TPR is computed over 21 positions and FPR over 29 (e.g., TPR \(= 95.2\% = 20/21\)); for recurring drift, the post-onset region alternates between active and null blocks of five buckets. Both sides are capped positionally rather than sampled at random: the Spark tests compare each full bucket (\({\approx}2.75\)M rows) with the first 100,000 rows of a separate reference table, and the driver-side tests compare the first 20,000 rows of each bucket with the first 10,000 reference rows. To bound the per-position cost, the scans use lighter settings than the Failing Loudly runs of Table~\ref{tab:tests-overview}: RFF-MMD with \(D = 200\) random features and no permutation null (decision by \(\tau_{\mathrm{RFF}}\)), SWD with 25 projections and 100 quantiles, MMD-RBF with 200 permutations on 3,000-row subsamples, and projected KS with 64 dense Gaussian projections. All results come from a single injection run per scenario; we report Wilson 95\% intervals for TPR and FPR and moving-block bootstrap intervals (block length 5, 2,000 resamples) for \(r\). Because positions share one reference and are not independent (Section~\ref{subsec3-3}), the Wilson intervals are approximate; the block bootstrap accounts for short-range dependence between neighboring positions.

A binary decision requires a threshold, and choosing one by inspecting the trajectories would leak into the reported metrics. For RFF-MMD we therefore calibrate on null data only: the 29 pre-drift positions of the sudden--strong scenario give an empirical 99th percentile of \(2.93 \times 10^{-3}\) (mean \(1.95 \times 10^{-4}\), std \(7.16 \times 10^{-4}\)), which we round up to \(\tau_{\mathrm{RFF}} = 3 \times 10^{-3}\) and apply unchanged to all eight scenarios. The pre-drift statistic is homogeneous across scenarios (means in \([1.71, 2.93] \times 10^{-4}\), standard deviations in \([6.5, 7.9] \times 10^{-4}\), with sudden--strong inside both ranges), although matching means and spreads do not establish matching upper tails, which a full cross-scenario tail check would require. Three caveats bound this choice. First, 29 samples that share one reference sample support an empirical high-quantile estimate, not a \(1\%\) size guarantee. Second, the quantile is dominated by the position-25 outlier (\(\approx 20\times\) the pre-drift mean; Section~\ref{subsubsec6-3-6}), which we retain; recalibrating without it is left to future work. Third, the pre-drift positions of sudden--strong also enter the headline FPR, which is therefore partially in-sample, whereas its TPR positions lie outside the calibration draw. The remaining tests keep their implementation thresholds (SWD \(\tau_{\mathrm{SW}} = 0.05\); per-dim KS Bonferroni \(\alpha_0 / d\)); their calibration behavior is discussed in Sections~\ref{subsubsec6-3-7} and~\ref{sec7}.

\subsubsection{Overall performance}\label{subsubsec6-3-3}

Table~\ref{tab:trendyol-grand-summary} summarizes the five tests averaged over all 8 scenarios. The per-bucket ground truth is \(\bar m_b\), the mean of \(m\) over the rows of bucket \(b\) (Eq.~\ref{eq:expected-magnitude}); we partition positions into \emph{injection-null} \(\mathcal{P}_0 = \{b : \bar m_b = 0\}\) and \emph{injection-active} \(\mathcal{P}_1 = \{b : \bar m_b > 0\}\), chosen in place of a chronological pre/post-drift split so that recurring-drift return-to-A blocks (post-onset positions with \(\bar m_b = 0\)) are correctly classified. FPR, TPR, ratio, and \(r\) are then defined on this partition per the Table~\ref{tab:trendyol-grand-summary} caption.

Two tests stand out. Spark RFF-MMD, under the calibrated threshold of Section~\ref{subsubsec6-3-2}, produces the most consistent discrimination: mean \(r = 0.427\), FPR \(= 3.2\%\), ratio \(= 8.41\) (injection-active statistic reaches \({\approx}8\times\) injection-null). Spark SWD produces a similar correlation (\(r = 0.422\)) but higher FPR (10.0\%) under its fixed threshold (\(\tau_\mathrm{SW} = 0.05\)). These grand means, however, obscure a critical stratification across the severity--scope regime: both tests produce defensible strong-regime results but fail to discriminate in the weak regime (Section~\ref{subsubsec6-3-4}). The remaining three tests (per-dim KS, MMD-RBF, projected KS) exhibit systematic failures whose mechanisms are summarized in Section~\ref{subsubsec6-3-7} and detailed in Section~\ref{sec7}.

\begin{table}[!htbp]
\caption{Sliding-scan detection metrics on the Trendyol collection-ranking feature table averaged over all 8 test scenarios (4 drift types \(\times\) 2 severity regimes). Grand means include the weak regime, where both RFF-MMD and SWD fail to discriminate; see Table~\ref{tab:trendyol-per-regime} for the per-regime breakdown.}\label{tab:trendyol-grand-summary}
\begin{tabular}{@{}lrrrrr@{}}
\toprule
\textbf{Test} & \textbf{TPR (\%)} & \textbf{FPR (\%)} & \textbf{Ratio} & \textbf{\(r\)} & \textbf{Time (s)} \\ \midrule
\textbf{Per-dim KS}      & 100.0 & 100.0 &  0.91 & \(-0.085\) &  0.07 \\
\textbf{MMD-RBF}         & 100.0 &  98.4 &  1.99 & \(-0.022\) & 36.20 \\
\textbf{Projected KS}    & 100.0 & 100.0 &  1.25 & \(+0.109\) &  0.63 \\
\textbf{Spark RFF-MMD}   &  40.2 &   3.2 &  8.41 & \(+0.427\) &  2.68 \\
\textbf{Spark SWD}       &  48.2 &  10.0 &  1.84 & \(+0.422\) &  5.50 \\
\botrule
\end{tabular}
\footnotetext{TPR = rejection rate over injection-active positions (\(\mathcal{P}_1 = \{b : \bar m_b > 0\}\)); FPR = \emph{injection-relative} rejection rate over injection-null positions (\(\mathcal{P}_0 = \{b : \bar m_b = 0\}\); this partition rather than a chronological pre/post-drift split is used because it correctly handles recurring drift). The injection-relative FPR is not the classical statistical Type-I error, since injection-null positions may still carry natural production-side covariate heterogeneity as documented in Section~\ref{subsubsec6-3-6}; Ratio = injection-active mean statistic divided by injection-null mean statistic; \(r\) = Pearson correlation between test statistic and per-bucket expected drift magnitude \(\bar m_b\) (Eq.~\ref{eq:expected-magnitude}); Time = median compute time per position on the Spark configuration described in Section~\ref{subsec4-2}. Spark RFF-MMD uses the calibrated threshold \(3 \times 10^{-3}\) (Section~\ref{subsubsec6-3-2}); the other tests use their implementation defaults. The three tests reporting near-100\% FPR and TPR simultaneously do not perform meaningful binary discrimination at their default thresholds; their failure mechanisms are discussed in Section~\ref{subsubsec6-3-7}.}
\end{table}

\subsubsection{Performance by drift type and severity regime}\label{subsubsec6-3-4}

Table~\ref{tab:trendyol-per-regime} reports strong- and weak-regime results side by side for each drift type, with the corresponding scenario-level scale in Table~\ref{tab:scenario-config}. \emph{The strong regime yields consistent detection across drift types, whereas the weak regime yields no reliable discrimination for either test.}

\begin{table}[!htbp]
\caption{Strong- vs weak-regime detection performance by drift type on the Trendyol collection-ranking feature table. RFF-MMD and SWD are the two tests capable of producing meaningful discrimination; the remaining three (per-dim KS, MMD-RBF, projected KS) exhibit systematic failures at all regimes (see Section~\ref{subsubsec6-3-7}).}\label{tab:trendyol-per-regime}
\centering
\footnotesize
\setlength{\tabcolsep}{4pt}
\begin{tabular*}{\textwidth}{@{\extracolsep\fill}ll rr rr rr rr @{}}
\toprule
& & \multicolumn{2}{c}{\textbf{Sudden}} & \multicolumn{2}{c}{\textbf{Gradual}} & \multicolumn{2}{c}{\textbf{Incremental}} & \multicolumn{2}{c}{\textbf{Recurring}} \\
\cmidrule(lr){3-4}\cmidrule(lr){5-6}\cmidrule(lr){7-8}\cmidrule(lr){9-10}
\textbf{Test} & \textbf{Metric} & \textbf{Strong} & \textbf{Weak} & \textbf{Strong} & \textbf{Weak} & \textbf{Strong} & \textbf{Weak} & \textbf{Strong} & \textbf{Weak} \\ \midrule
\multirow{3}{*}{\textbf{RFF-MMD}}
     & \(r\)   & \(+0.948\) & \(-0.094\) & \(+0.944\) & \(-0.081\) & \(+0.942\) & \(-0.083\) & \(+0.926\) & \(-0.084\) \\
     & TPR   & 95.2 & 0.0 & 71.4 & 0.0 & 71.4 & 0.0 & 83.3 & 0.0 \\
     & FPR   &  3.4 & 3.4 &  3.4 & 3.4 &  3.4 & 3.4 &  2.6 & 2.6 \\
\midrule
\multirow{3}{*}{\textbf{SWD}}
     & \(r\)   & \(+0.844\) & \(+0.033\) & \(+0.844\) & \(+0.095\) & \(+0.820\) & \(+0.041\) & \(+0.767\) & \(-0.069\) \\
     & TPR   & 95.2 &  9.5 & 76.2 & 14.3 & 76.2 & 14.3 & 83.3 & 16.7 \\
     & FPR   & 10.3 & 10.3 &  6.9 & 10.3 & 10.3 & 10.3 & 13.2 &  7.9 \\
\botrule
\end{tabular*}
\footnotetext{Metric definitions (\(r\), TPR, FPR and the \(\mathcal{P}_1\)/\(\mathcal{P}_0\) partition) as in Table~\ref{tab:trendyol-grand-summary}. Each cell is a single scenario; TPR and FPR in \%. RFF-MMD uses the calibrated threshold \(3 \times 10^{-3}\) (Section~\ref{subsubsec6-3-2}); SWD uses the fixed implementation threshold \(\tau_\mathrm{SW} = 0.05\). With 21 active and 29 null positions per scenario (12 and 38 for recurring drift), Wilson 95\% intervals are wide, e.g., \(20/21 = 95.2\%\) [77.3, 99.2], \(15/21 = 71.4\%\) [50.0, 86.2], \(1/29 = 3.4\%\) [0.6, 17.2], and \(3/29 = 10.3\%\) [3.6, 26.4].}
\end{table}

In the strong regime, RFF-MMD reaches \(r \in [0.93, 0.95]\). The range stays tight because \(m\) interpolates through gradual and incremental transitions at the intensity scale the statistic tracks, whereas an indicator-based target would collapse to a step function. Its TPR is 95\% for sudden, 71\% for gradual and incremental, and 83\% for recurring drift, while FPR stays within \([2.6, 3.4]\%\), so the single calibrated threshold generalizes across drift types. SWD follows the same pattern at a lower level (\(r \in [0.77, 0.84]\), TPR \(\in [76.2, 95.2]\%\)) but with an FPR of 6.9--13.2\%, because its fixed threshold \(\tau_\mathrm{SW} = 0.05\) is not null-calibrated (Section~\ref{sec7}). These per-scenario differences, across drift types and between the two tests, lie within wide uncertainty bands: the Wilson intervals in Table~\ref{tab:trendyol-per-regime} span roughly 17--36 percentage points, and the bootstrap intervals for \(r\) overlap between RFF-MMD (lower bounds 0.63--0.80) and SWD (lower bounds 0.43--0.66).

In the weak regime, RFF-MMD detects nothing (TPR \(= 0\%\), \(r \in [-0.094, -0.081]\)) and SWD reaches a TPR of only \(9.5\)--\(16.7\%\) with \(r \approx 0\). At realized-flip fractions of $0.29$--\(0.57\%\), a shift of this size on the 11-dimensional vector cannot be distinguished from the injection-null baseline under the sample sizes and thresholds used here. The two regimes thus bracket a failure and a success region without locating the power transition between them; an intensity-grid power analysis (e.g.\ \(\{0.5, 1, 2, 3, 5, 7.5, 10\}\%\) with repeated seeds) is left to future work, and the weak-regime results mark an empirical failure point for this configuration rather than a fundamental sensitivity bound.

\begin{figure*}[p]
\centering
\includegraphics[width=\textwidth,height=0.78\textheight,keepaspectratio]{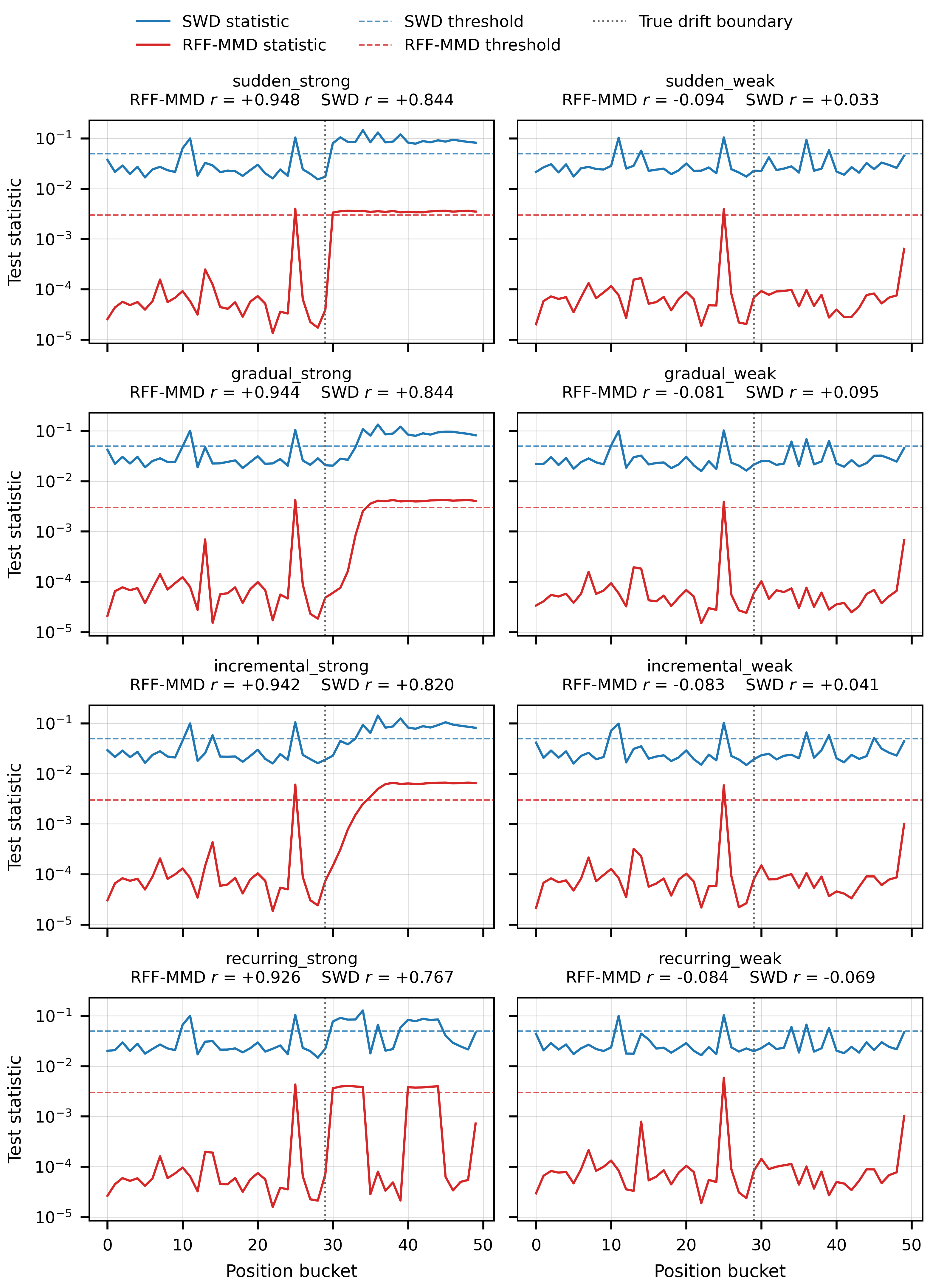}
\caption{Sliding-scan trajectories of Spark RFF-MMD (red) and Sliced Wasserstein Distance (blue) across 8 scenarios. Rows correspond to the four drift types; the left column is the strong regime and the right column the weak regime. Solid lines are the test statistic on a shared log y-axis; dashed horizontal lines are the drift threshold of each test in the matching color (RFF-MMD: \(3 \times 10^{-3}\), calibrated per Section~\ref{subsubsec6-3-2}; SWD: \(5 \times 10^{-2}\), implementation default). The dotted vertical line marks the initial drift onset (for recurring drift, subsequent returns between concepts are additional true transition points not marked). Per-panel Pearson \(r\) is shown in each subplot title.\label{fig:trendyol-trajectory-rff-swd}}
\end{figure*}

\subsubsection{Trajectory analysis}\label{subsubsec6-3-5}

Figure~\ref{fig:trendyol-trajectory-rff-swd} shows per-bucket sliding-scan trajectories of Spark RFF-MMD (red) and Sliced Wasserstein Distance (blue) across all four drift types under both strong and weak severity regimes. In the strong regime, both tests exhibit the expected schedule-specific patterns: a sharp single step at bucket 29 for sudden drift; a smooth ramp for gradual and incremental; alternating on/off cycles for recurring. In the weak regime, the trajectories remain indistinguishable from the injection-null baseline.

\subsubsection{Position-25 anomaly}\label{subsubsec6-3-6}

Position 25 exhibits a systematic detection spike across all tests unrelated to injected drift, attributable to natural source-table heterogeneity (multi-day bucket boundary crossing daytime traffic patterns); this is a documented finding of the study, with detailed analysis in Appendix~\ref{secA4}.

\subsubsection{Failure modes: per-dim KS, MMD-RBF, and projected KS}\label{subsubsec6-3-7}

Three tests in Table~\ref{tab:trendyol-grand-summary} fail to produce a meaningful correlation between statistic and ground truth and/or exhibit near-100\% FPR; brief diagnoses follow, with full mechanistic discussion in Section~\ref{sec7}.

\begin{itemize}
    \item \textbf{Per-dim KS} (\(r = -0.085\), FPR \(=\) TPR \(= 100\%\)): The Bonferroni-corrected max-KS statistic is dominated by ID-like columns (seed-content and owner identifiers, with \(\approx 5.9\)M and $183$k distinct values respectively) under the asymmetric reference/current sampling design used here (reference: first 10,000 rows of the reference table; bucket: first 20,000 rows of each stream-ordered bucket). The two sampling procedures observe systematically different empirical distributions on these columns and produce \(D \approx 0.20\) independent of drift, dominating the true drift signal on the \(y'\) dimension (\(D \approx 0.06\)). Boolean detection fires at every position; statistic magnitude shows no drift-tracking pattern. Disentangling high-cardinality per se from sampling asymmetry requires an ablation (matched sampling, ID-column removal) identified as future work in Section~\ref{sec7}.

    \item \textbf{MMD-RBF} (\(r = -0.022\), FPR \(= 98.4\%\)): The kernel bandwidth is estimated by the median heuristic on 3\,000-row driver subsamples drawn separately for the reference and each bucket; because samples come from different regions of the stream, bandwidth estimates differ by a few percent across buckets, and we hypothesise that this bandwidth mismatch --- interacting with subsampling and the fixed threshold --- decorrelates the MMD statistic from the injected drift signal. The same implementation reproduces published detection accuracies on Failing Loudly (Section~\ref{subsec6-2}), suggesting the failure is scale-and-sampling-specific; a full mechanistic analysis at Trendyol scale is deferred to future work.

    \item \textbf{Projected KS} (\(r = +0.109\), FPR \(=\) TPR \(= 100\%\)): A low but non-zero correlation indicates partial signal capture on random-projected features, but always-reject behavior under the fixed threshold makes the test unusable for production.
\end{itemize}

\begin{figure*}[!htbp]
\centering
\includegraphics[width=\textwidth]{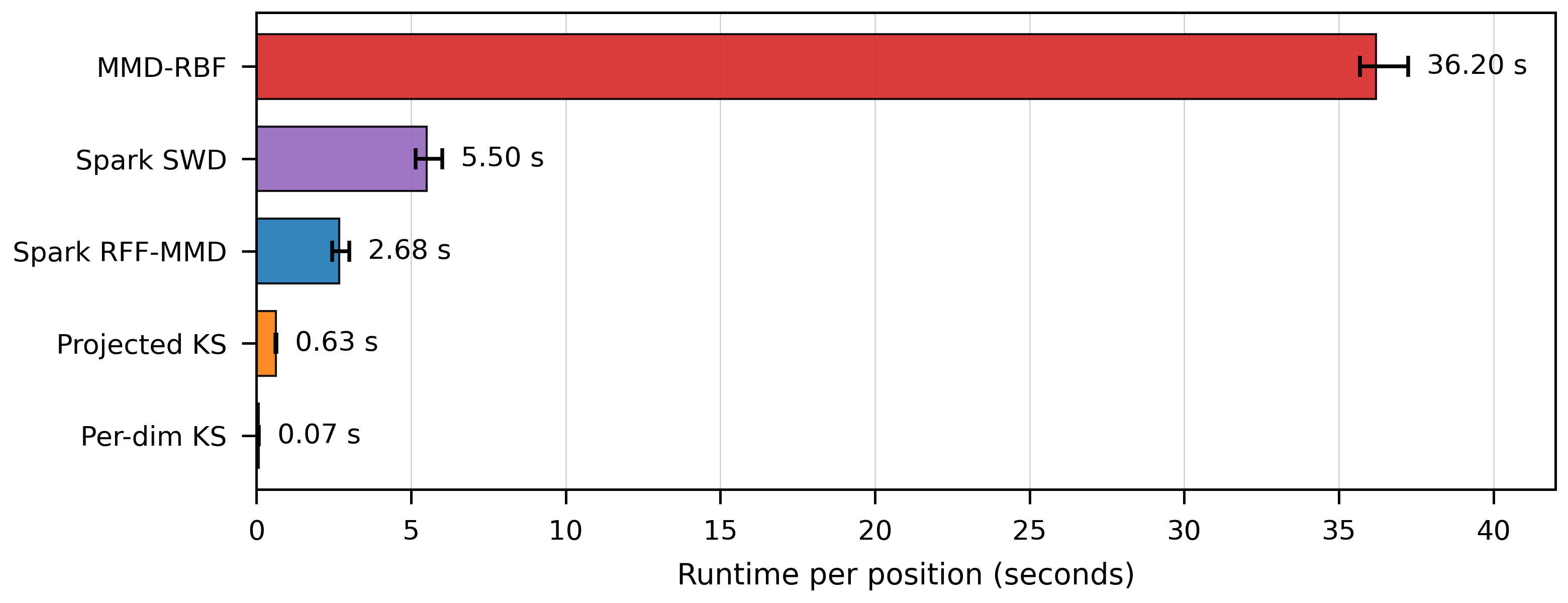}
\caption{Per-position runtime aggregated across all 8 scenarios (\(n = 400\) measurements per test: 50 positions \(\times\) 8 scenarios). Bars are medians; whiskers span the interquartile range. Each test is assigned a distinct color that is reused consistently in Figure~\ref{fig:trendyol-runtime-boxplot}. MMD-RBF dominates the range at $\sim$36 seconds per position, roughly 6.5$\times$ slower than the next test (Sliced Wasserstein Distance) and roughly 480\(\times\) slower than per-dim KS.\label{fig:trendyol-runtime-bar}}
\end{figure*}

\begin{figure*}[!htbp]
\centering
\includegraphics[width=\textwidth]{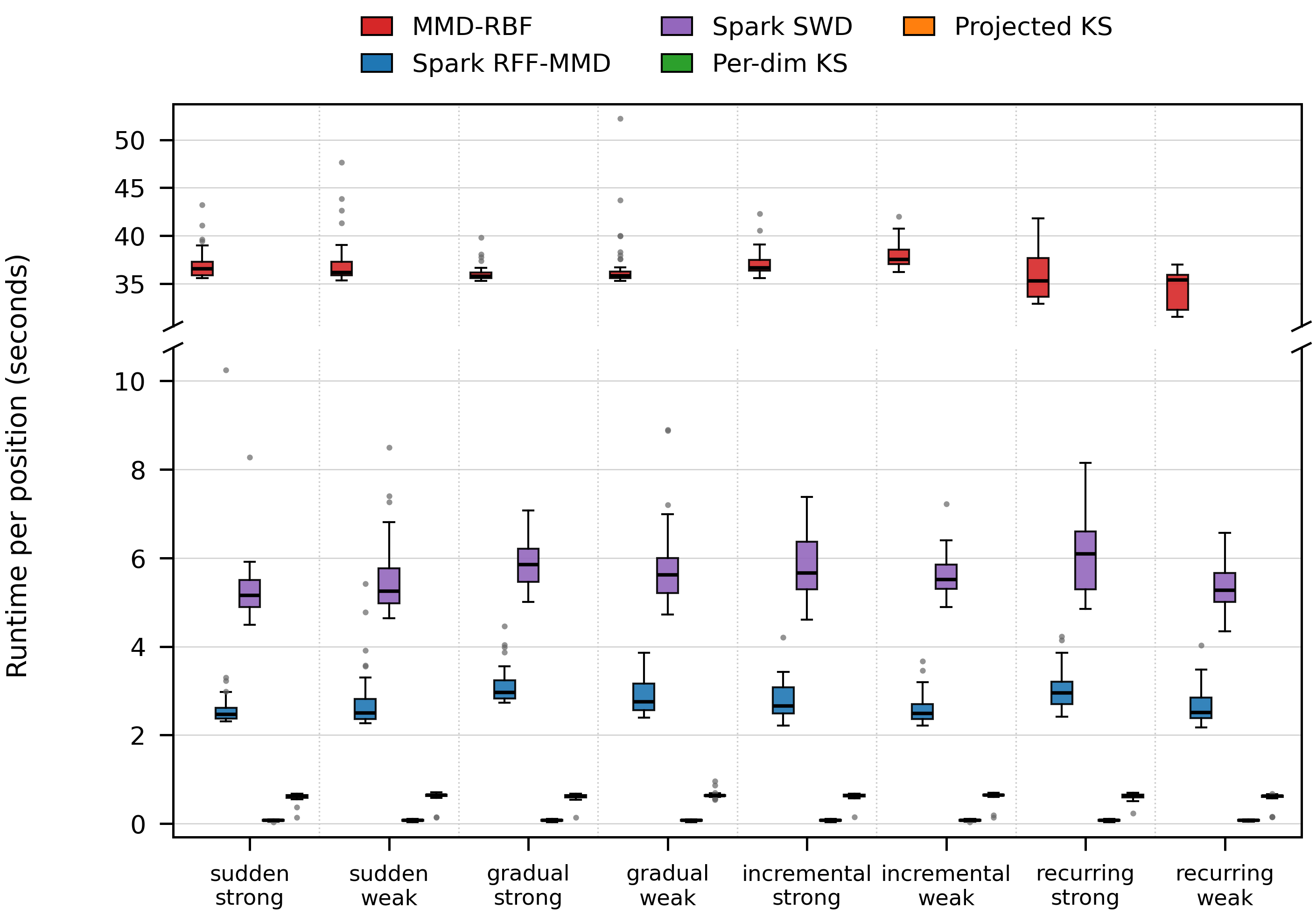}
\caption{Per-position runtime distribution disaggregated by scenario. Each box aggregates \(n = 50\) per-position measurements from one scenario. Colors match Figure~\ref{fig:trendyol-runtime-bar}. The y-axis is broken to accommodate MMD-RBF's range (\({\sim}31\)--53 s, top panel) while preserving resolution on the four smaller-scale tests (0--10.5 s, bottom panel). Distributions are similar across scenarios, confirming that runtime is a function of test complexity and execution mode, independent of drift structure or severity regime.\label{fig:trendyol-runtime-boxplot}}
\end{figure*}

\subsubsection{Runtime profile}\label{subsubsec6-3-8}

Figures~\ref{fig:trendyol-runtime-bar} (aggregate median with interquartile-range whiskers) and~\ref{fig:trendyol-runtime-boxplot} (disaggregated by scenario) report per-position compute cost for the five tests: runtime is stable across drift types and severity regimes, confirming it is a property of test complexity and execution mode rather than of the injected drift structure.

\FloatBarrier 
\section{Discussion and Implications for Big Data Systems}\label{sec7}

The three benchmarks are complementary rather than interchangeable: no single one can characterize a drift detector, because each answers a different question about what the tests actually \emph{see}. The Harvard Dataverse streams probe operational structure: with \(P(X)\) stationary and only \(P(Y \mid X)\) changing, multi-column tests detect drift because the class column is part of the test vector (Sections~\ref{subsec5-4} and~\ref{subsec6-1}). Failing Loudly probes implementation consistency, providing a numerical reproduction target (MAE \(\in [0.030, 0.053]\); Section~\ref{subsec6-2}). Neither establishes transfer to production scale. The Trendyol benchmark supplies that third axis: the tests continue to produce signal on \({\sim}137.5\) million rows and, after the pre-check of Section~\ref{app-precheck}, on an 11-dimensional feature vector, with conclusions bounded by a single collection-ranking table. The contribution of this study is therefore not a new method but a validation topology.

Univariate marginal tests (chi-square, PSI, Jensen--Shannon, univariate KS, Wasserstein) dominate production monitoring because they are cheap and easy to report, yet our results show that marginal monitoring fails in three distinct ways. \emph{First}, per-dimension tests cannot see changes in the dependence structure between features; detecting correlation changes under fixed marginals requires a multivariate statistic. Controlled experiments show that feature-wise KS performs close to chance when drift affects only the correlation between features, whereas kernel-based tests do not \citep{hinder2024survey}, a result that PSI-based monitoring proposals routinely overlook. \emph{Second}, at scale, per-dimension KS can be captured by the wrong columns. On the pre-check-narrowed Trendyol vector, Bonferroni-corrected per-dim KS rejects at every position (FPR \(=\) TPR \(= 100\%\), \(r = -0.085\)). The cause is not the classical \(n \to \infty\) \(p\)-value collapse: two ID-like columns (seed-content and owner identifiers, with \(\approx 5.9\)M and \(\approx 183\)k distinct values) differ systematically between the positionally capped reference sample and the stream-ordered current buckets, producing \(D \approx 0.20\) regardless of drift and swamping the true signal on the \(y'\) dimension (\(D \approx 0.06\)). Whether high cardinality or the asymmetric sampling design is responsible requires an ablation (matched sampling and, separately, ID-column removal) that we leave to future work; in the interim, either intervention is a practical safeguard, whereas conformal or bootstrap threshold calibration addresses neither mechanism directly. \emph{Third}, the ``marginal blindness'' argument needs refinement on the label axis. On SINE and MIXED, per-dim KS reaches 0\% recall while projected KS and MMD-RBF reach 100\% (Section~\ref{subsec6-1}); but because the multi-column vector includes the one-hot class column, that success may stem from the class marginal, from \(X\)--\(Y\) dependence, or both, and only an X-only / Y-only / \((X, Y)\) / shuffled-Y ablation would separate them. The defensible claim is narrower: Bonferroni-corrected per-dimension tests lose \emph{any} meaningful signal, including the label marginal, whereas multi-column tests observe a joint signal via projection or kernel. When an outcome \(Y\) or prediction \(\hat{Y}\) channel exists, adding it to the test vector strengthens detection, as a layer on top of, not a replacement for, monitoring \(P(X)\). Because the Trendyol benchmark uses a constructed \(Y_{\text{synth}}\) in place of such a channel, its results describe \((X, Y_{\text{synth}})\) joint-distribution detection rather than pure \(P(X)\) detection.

Random Fourier Features are what make kernel testing viable at scale. Exact MMD-RBF costs \(O(N^2 d)\) and is impractical at the Trendyol reference size (\(N = 100\,000\)). Their approximation \citep{rahimi2007random} promises linear time, but whether they preserve the kernel statistic's signal at production scale was unknown. Our results answer on both fronts. Under the Failing Loudly protocol, RFF-MMD (\(D = 500\)) matches MMD-RBF's detection accuracy with a size-controlled permutation \(p\)-value (Appendix~\ref{app-subsubsec6-4-1}). In the Trendyol strong regime it reaches, averaged over drift types, \(r = 0.940\), TPR \(= 80.4\%\), and FPR \(= 3.2\%\) under a threshold calibrated on injection-null pre-drift positions (Section~\ref{subsubsec6-3-2}), at a median cost of $2.68$~s per position. Distributed RFF-MMD is therefore a practical building block for monitoring streams of millions of rows. The weak regime marks its limit: at realized-flip fractions of \(0.29\)--\(0.57\%\), TPR falls to \(0\%\) and \(r \approx -0.09\). Locating the power transition between the two regimes requires an intensity-grid sweep with repeated seeds.

Finally, a useful signal is not a calibrated test, as the Sliced Wasserstein Distance (SWD) illustrates. Under Failing Loudly with the fixed threshold \(\tau_{\mathrm{SW}} = 0.05\), its false-positive rate never approaches the nominal level across 60 no-shift trials and the Wilson 95\% interval never contains $0.05$ (Appendix~\ref{app-subsubsec6-4-1}): the present implementation is not size-controlled. Yet in the Trendyol strong regime it tracks drift nearly as well as RFF-MMD (mean \(r = 0.819\), TPR \(= 82.7\%\)), at the cost of FPR \(\approx 10\%\) under the uncalibrated threshold, and it shares RFF-MMD's weak-regime failure. This behavior is consistent with a documented limitation of uniformly sampled slicing: when the displacement between two distributions concentrates in a low-dimensional subspace, the probability that a random projection direction is nearly orthogonal to it grows exponentially in \(d\), so the estimator underestimates the true distance \citep{kolouri2019generalized,nguyen2021distributional}. Adaptive slicing variants (max-SW, max-generalized SW, distributional SW), a permutation null, and the null calibration used for RFF-MMD are natural remedies that lie outside the scope of this paper. The broader lesson for production systems is that an unsupervised distance-based signal can serve as an observational alert, but operational decisions require a size-controlled test; the conformal calibration of \citet{leong2026conformal} may bridge the two but has not yet been validated at production scale.

\section{Conclusion}\label{sec8}

Multivariate drift detection works at production scale, but only when test choice, calibration, and sampling design are treated as part of the method. We evaluated five multi-column two-sample tests across three complementary layers---controlled concept-drift streams (Harvard Dataverse), a reproduced image-shift benchmark (Failing Loudly), and synthetic concept drift injected into the 137.5-million-row Trendyol collection-ranking feature table---and draw four conclusions.

\emph{First, distributed RFF-MMD is the strongest candidate for production monitoring.} In the strong regime it tracks expected drift magnitude with, averaged over drift types, \(r = 0.940\), a true positive rate of \(80.4\%\), and a false positive rate of \(3.2\%\) under a null-calibrated threshold, at $2.68$\,s per position; across the eight scenarios, the five tests together required about 5.1 hours of cumulative compute, of which RFF-MMD accounted for about 19 minutes. Random Fourier Features thus turn kernel testing from an \(O(N^2)\)-memory method into a practical Spark workload.

\emph{Second, a drift signal is not a decision rule.} The Sliced Wasserstein Distance tracks drift nearly as well (TPR \(= 82.7\%\)) but, without calibration, flags about \(10\%\) of injection-null positions, while per-dimension KS rejects everywhere, saturated by ID-like columns under asymmetric sampling. Calibration and sampling design, not the statistic alone, determine operational reliability.

\emph{Third, benchmark design is part of the result.} Sliding-window trajectories scored against row-level ground truth expose what binary recall hides, and pre-check and sanity-check protocols are what separate genuine drift from data-quality artifacts in production tables.

\emph{Fourth, every detector has a floor.} At realized-flip fractions of \(0.29\)--\(0.57\%\), neither RFF-MMD nor SWD separates drift from the null baseline; whether this is a fundamental sensitivity bound or a threshold that shifts with sample size and kernel choice requires intensity-grid power analysis with repeated seeds.

These results come from a single injection run on a single e-commerce table and describe joint \((X, Y_{\text{synth}})\) rather than pure \(P(X)\) detection. Extending them to other domains, attributing natural covariate heterogeneity separately from injected drift, and bringing adaptive slicing and conformal calibration to production scale are the natural next steps toward monitoring systems that fail loudly rather than silently.

\bibliography{sn-bibliography}

\section*{Abbreviations}
\noindent{\small
\begin{tabular}{@{}l>{\raggedright\arraybackslash}p{0.33\textwidth}@{\hspace{1em}}l>{\raggedright\arraybackslash}p{0.33\textwidth}@{}}
API & Application programming interface & PSI & Population Stability Index \\
FPR & False positive rate & RBF & Radial basis function \\
FWER & Family-wise error rate & RFF & Random Fourier Features \\
ID & Identifier & RFF-MMD & MMD with Random Fourier Features \\
JS & Jensen--Shannon & RKHS & Reproducing kernel Hilbert space \\
KL & Kullback--Leibler & SQL & Structured Query Language \\
KS & Kolmogorov--Smirnov & SRP & Sparse random projection \\
MAE & Mean absolute error & SW & Sliced Wasserstein \\
MMD & Maximum Mean Discrepancy & SWD & Sliced Wasserstein Distance \\
MMD-RBF & MMD with a radial basis function kernel & TPR & True positive rate \\
\end{tabular}}

\newpage
\begin{appendices}

\section{Notation}\label{secA1}

Table~\ref{tab:notation} lists the symbols used in the paper together with the place where each is defined.

\begin{table}[!htbp]
\caption{Notation used throughout the paper.}\label{tab:notation}
\centering
\footnotesize
\begin{tabular}{@{}>{\raggedright\arraybackslash}p{0.22\textwidth}>{\raggedright\arraybackslash}p{0.50\textwidth}>{\raggedright\arraybackslash}p{0.20\textwidth}@{}}
\toprule
\textbf{Symbol} & \textbf{Meaning} & \textbf{Defined in} \\ \midrule
\multicolumn{3}{@{}l}{\emph{Two-sample testing}} \\
\(\mathcal{D}_R, \mathcal{D}_C\) & Reference and current samples & Sec.~\ref{subsec3-1} \\
\(P_R, P_C\) & Distributions of the reference and current samples & Eq.~(\ref{eq:batch-null}) \\
\(z\); \(X, Y\) & Tested vector (\((X,Y)\) or \(X_k\)); features and label & Sec.~\ref{subsec3-1} \\
\(d\) & Feature dimensionality & Table~\ref{tab:tests-overview} \\
\(s\); \(p\) & Test statistic; \(p\)-value & Sec.~\ref{subsec3-1} \\
\(\psi\) & Decision rule & Sec.~\ref{subsec3-1} \\
\(\alpha_0\) & Significance level (0.05) & Sec.~\ref{subsec3-1} \\
\(\tau\); \(\tau_{\mathrm{RFF}}, \tau_{\mathrm{SW}}\) & Decision threshold; RFF-MMD (\(3\times10^{-3}\)) and SWD (0.05) thresholds & Secs.~\ref{subsec4-1}, \ref{subsubsec6-3-2} \\
\(\indicator[\cdot]\) & Indicator function & Sec.~\ref{subsec3-1} \\
\midrule
\multicolumn{3}{@{}l}{\emph{Sliding-window scan}} \\
\(N\) & Number of position buckets & Eq.~(\ref{eq:bucket-partition}) \\
\(b\); \(\mathcal{B}_b\) & Bucket index; current bucket & Eq.~(\ref{eq:bucket-partition}) \\
\(s_b\) & Test statistic at bucket \(b\) & Sec.~\ref{subsec3-2} \\
\(W\); \(\Delta\); \(t\) & Window size; stride; stream index (moving reference) & Sec.~\ref{subsec3-2} \\
\(\mathcal{P}_0, \mathcal{P}_1\) & Injection-null and injection-active positions & Sec.~\ref{subsubsec6-3-3} \\
\(r\) & Pearson correlation between \(s_b\) and \(\bar m_b\) & Sec.~\ref{subsubsec6-3-3} \\
\midrule
\multicolumn{3}{@{}l}{\emph{Drift injection and ground truth}} \\
\(Y_A, Y_B\) & Concept A and Concept B labeling rules & Eqs.~(\ref{eq:concept-a-rule}), (\ref{eq:concept-b-rule}) \\
\(\mathrm{sim}_{\mathrm{img}}, \mathrm{sim}_{\mathrm{txt}}\) & Image and text cosine similarity & App.~\ref{secA6} \\
\(y_A\); \(y'\); \(Y_{\text{synth}}\) & Pre-drift label; injected label; injected label variable & Sec.~\ref{subsec5-4} \\
\(\alpha\) & Per-row drift intensity & Sec.~\ref{subsec5-4} \\
\(\sigma\); \(\phi\) & Severity; scope & Sec.~\ref{subsec5-4} \\
\(s_0\); \(w\); \(\lambda\) & Drift start, transition width and recurrence period fractions & Alg.~\ref{alg:concept-drift-injection} \\
\(\delta\) & Concept A/B disagreement rate & Eq.~(\ref{eq:scenario-flip-rate}) \\
\(\bar\alpha_s\) & Mean drift intensity of scenario \(s\) & Eq.~(\ref{eq:scenario-flip-rate}) \\
\(e\); \(m\); \(f\) & Eligibility; expected drift magnitude; realized flip (per row) & Eqs.~(\ref{eq:eligible})--(\ref{eq:realized-flip}) \\
\(a\) & Coarse affected-row indicator & Alg.~\ref{alg:concept-drift-injection} \\
\(\bar m_b\); \(\bar f_b\) & Per-bucket means of \(m\) and \(f\) & Sec.~\ref{subsec5-4} \\
\midrule
\multicolumn{3}{@{}l}{\emph{Test configuration}} \\
\(D\) & RFF dimension (500; 200 in the Trendyol scans) & Table~\ref{tab:tests-overview} \\
\(B\) & Number of permutations & Table~\ref{tab:tests-overview} \\
\(K\) & Number of random projections (32 sparse; 64 dense in the Trendyol scans) & Table~\ref{tab:tests-execution} \\
\(P\) & Number of SWD projections (100; 25 in the Trendyol scans) & Table~\ref{tab:tests-overview} \\
\(n_{\mathrm{drv}}\) & Driver sample size (20,000) & Sec.~\ref{subsec4-2} \\
\botrule
\end{tabular}
\end{table}

\section{Pre-check and Sanity-check Procedures for Production Drift Injection}\label{secA2}

This appendix presents (i) the custom sliding-window scan framework used throughout Sections~\ref{subsec3-2}, \ref{subsec5-2}, and~\ref{subsec6-3} as pseudocode, and (ii) the pre-check and verification steps of the injection design defined in Section~\ref{subsec5-4}, likewise as procedural pseudo-code. The procedures are expressed in terms of a table and can be translated directly to any SQL-like query language or dataframe API (BigQuery, Spark, pandas).

\subsection{Pre-check protocol: active feature inventory}\label{app-precheck}

For any concept drift benchmark to produce meaningful results, the columns affected by injection must carry active variance --- a condition not always guaranteed in production tables where upstream aggregation may leave outputs empty, planned feature values may fail to materialize, or flag/count columns may be zeroed out. A recommended seven-criterion audit protocol (zero/near-zero variance; identifier semantics; cardinality ratio; monotonic structure; missingness; reference/current sampling compatibility; embedding array dimensionality) is listed here; Algorithm~\ref{alg:variance-inventory} implements its variance criterion, and the audit reported here scanned that criterion only, with the additional criteria consolidated post-hoc after the failure analysis of Section~\ref{sec7} showed they matter as much for benchmark-vector eligibility. In particular, the identifier-semantics criterion was not used to filter the benchmark vector: the three identifier columns (seed-content, owner, and channel identifiers, two of them with high cardinality) were retained by design, because such columns are routinely present in production feature tables and a monitoring system must behave sensibly in their presence. Before testing, every vector component is z-scored with the reference statistics, and missing values are imputed with the reference mean.

The variance scan of the 164-column source table returned three groups: 148 of 158 scalar numeric columns held a single value (0) across all rows --- these included the natural injection candidates (action flag, user count, cart count, and collection-click flag) and were entirely dormant; 10 scalar columns carried genuine variance (4 metadata scalars --- seed-content identifier, owner identifier, channel identifier, and collection size --- and 6 similarity metrics: image, tag, and text cosine similarity plus the corresponding matched-content counts); and 6 embedding arrays (three collection, three seed; dimensions 128, 256, 415) which were audited separately and carried variance. Table~\ref{tab:feature-glossary} lists the resulting column groups and maps the descriptive names and symbols used in the text to the original column names.

This finding had three direct consequences for the injection design: (i) dormant columns cannot serve as injection targets (a Bernoulli flip on a constant-zero column is undefined); (ii) dormant columns must be excluded from the benchmark vector or they dilute the drift signal and disturb kernel bandwidth estimation; and (iii) the synthetic label must be defined over live-variance features only, which we do over the similarity metrics (\S\ref{subsec5-4}).

\begin{table}[!htbp]
\caption{Feature and injection-variable glossary for the Trendyol benchmark. Descriptive names and symbols used in the text are mapped to the original column names of the source table.}\label{tab:feature-glossary}
\centering
\footnotesize
\begin{tabular}{@{}>{\raggedright\arraybackslash}p{0.26\textwidth}>{\raggedright\arraybackslash}p{0.12\textwidth}>{\raggedright\arraybackslash}p{0.22\textwidth}>{\raggedright\arraybackslash}p{0.28\textwidth}@{}}
\toprule
\textbf{Name in Text} & \textbf{Symbol} & \textbf{Role} & \textbf{Original Column} \\ \midrule
\multicolumn{4}{@{}l}{\emph{Source-table columns}} \\
\textbf{Seed-content identifier} & -- & Metadata (vector) & \texttt{seed} \\
\textbf{Owner identifier} & -- & Metadata (vector) & \texttt{ownerid} \\
\textbf{Channel identifier} & -- & Metadata (vector) & \texttt{channelId} \\
\textbf{Collection size} & -- & Metadata (vector) & \texttt{content\_\allowbreak cnt} \\
\textbf{Image cosine similarity} & \(\mathrm{sim}_{\mathrm{img}}\) & Similarity (vector); Concept A & \texttt{image\_\allowbreak cosine\_\allowbreak similarity} \\
\textbf{Text cosine similarity} & \(\mathrm{sim}_{\mathrm{txt}}\) & Similarity (vector); Concept B & \texttt{text\_\allowbreak cosine\_\allowbreak similarity} \\
\textbf{Tag cosine similarity} & -- & Similarity (vector) & \texttt{tag\_\allowbreak cosine\_\allowbreak similarity} \\
\textbf{Matched-content counts} & -- & Similarity (vector) & \texttt{*\_\allowbreak matched\_\allowbreak contents\_\allowbreak count} \\
\textbf{Collection identifier} & -- & Row key (hashing) & \texttt{collectionid} \\
\textbf{Recommendation source} & -- & Not in vector & \texttt{source} \\
\textbf{Action, user-count, cart-count and collection-click fields} & -- & Dormant & \texttt{action\_\allowbreak flag}, \texttt{user\_\allowbreak count}, \texttt{cart\_\allowbreak count}, \texttt{collection\_\allowbreak click\_\allowbreak flag} \\
\midrule
\multicolumn{4}{@{}l}{\emph{Columns written by the injection (Algorithm~\ref{alg:concept-drift-injection})}} \\
\textbf{Injected label} & \(y'\), \(Y_{\text{synth}}\) & Label (vector) & \texttt{synthetic\_\allowbreak y} \\
\textbf{Pre-drift label} & \(y_A\) & Verification & \texttt{original\_\allowbreak synthetic\_\allowbreak y} \\
\textbf{Drift intensity} & \(\alpha\) & Ground truth & \texttt{drift\_\allowbreak alpha} \\
\textbf{Eligibility indicator} & \(e\) & Ground truth (Eq.~\ref{eq:eligible}) & \texttt{eligible} \\
\textbf{Expected drift magnitude} & \(m\) & Ground truth (Eq.~\ref{eq:expected-magnitude}) & \texttt{expected\_\allowbreak drift\_\allowbreak magnitude} \\
\textbf{Realized-flip indicator} & \(f\) & Ground truth (Eq.~\ref{eq:realized-flip}) & \texttt{realized\_\allowbreak flip} \\
\textbf{Affected indicator} & \(a\) & Ground truth (coarse) & \texttt{is\_\allowbreak row\_\allowbreak affected} \\
\textbf{Stream index} & \(i\) & Row ordering & \texttt{stream\_\allowbreak index} \\
\botrule
\end{tabular}
\end{table}

\subsection{Sanity-check protocol: post-injection verification}\label{app-sanity}

Each injected table is verified in four parts before it is scanned (pseudocode in Section~\ref{app-pseudocode}): (i) the \emph{flip-rate check} (Algorithm~\ref{alg:flip-rate-check}) compares each scenario's empirical flip rate with the prediction of Eq.~(\ref{eq:scenario-flip-rate}); (ii) the \emph{zero-flip check} (Algorithm~\ref{alg:zero-flip-check}) confirms that \(y' = y_A\) for every row with \(\alpha = 0\); (iii) the \emph{covariate side-effect check} (Algorithm~\ref{alg:covariate-stationarity}) compares summary statistics of each feature before and after the drift onset; and (iv) the \emph{ground-truth consistency check} (Algorithm~\ref{alg:ground-truth-check}) aggregates \(m\) and \(f\) per bucket and confirms that the two targets agree.

\subsection{Pseudocode: pre-check and sanity-check procedures}\label{app-pseudocode}

\begin{algorithm}[!ht]
\caption{Sliding-Window Changepoint Detection}\label{alg:sliding-scan}
\begin{algorithmic}[1]
\Require Reference set \(\mathcal{D}_R\); stream-ordered current dataset \(\mathcal{D}_C\); number of positions \(N\); two-sample tests \(\mathcal{T} = \{T_1, \ldots, T_k\}\) with decision rules \(\{\psi_j\}_{j=1}^k\) (Section~\ref{subsec3-1}).
\Ensure Trajectory matrix \(\mathbf{S} \in \mathbb{R}^{N \times k}\); decision matrix \(\mathbf{D} \in \{0,1\}^{N \times k}\); optional ground-truth vector \(\mathbf{g} \in [0,1]^N\).
\State Partition \(\mathcal{D}_C\) into \(\{\mathcal{B}_0, \ldots, \mathcal{B}_{N-1}\}\) per Equation~(\ref{eq:bucket-partition})
\ForAll{\(b \in \{0, 1, \ldots, N-1\}\)}
    \If{\(\mathcal{B}_b\) carries the injection ground truth \(\alpha\) and \(m\)}
        \State \(\mathbf{g}[b] \gets \frac{1}{|\mathcal{B}_b|} \sum_{z \in \mathcal{B}_b} m(z)\) \Comment{ground-truth aggregation, Section~\ref{subsec3-3}}
    \EndIf
    \ForAll{\(T_j \in \mathcal{T}\)}
        \State \((s_b^{(j)},\, p_b^{(j)}) \gets T_j(\mathcal{D}_R,\, \mathcal{B}_b)\) \Comment{two-sample test, Section~\ref{sec4}}
        \State \(\mathbf{S}[b, j] \gets s_b^{(j)}\)
        \State \(\mathbf{D}[b, j] \gets \psi_j\bigl(s_b^{(j)}, p_b^{(j)}\bigr)\) \Comment{\(p\)-value or threshold rule}
    \EndFor
\EndFor
\State \Return \((\mathbf{S},\, \mathbf{D},\, \mathbf{g})\)
\end{algorithmic}
\end{algorithm}

\begin{algorithm}[!ht]
\caption{Active feature inventory (variance scan)}\label{alg:variance-inventory}
\begin{algorithmic}[1]
\Require Source table \(T\); list of scalar numeric columns \(C = \{c_1, \ldots, c_N\}\).
\Ensure For each column, the triple \((\mathrm{distinct}(c), \mathrm{mean}(c), \mathrm{sd}(c))\) of distinct-value count, mean and standard deviation; partition of \(C\) into \emph{alive} and \emph{dormant} columns.
\State Compute \((\mathrm{distinct}(c_i), \mathrm{mean}(c_i), \mathrm{sd}(c_i))\) for all \(c_i \in C\) in a single scan of \(T\)
\ForAll{\(c_i \in C\)}
    \If{\(\mathrm{distinct}(c_i) \leq 1\)}
        \State Add \(c_i\) to the \textsc{Dormant} list
    \Else
        \State Add \(c_i\) to the \textsc{Alive} list; record \((\mathrm{mean}(c_i), \mathrm{sd}(c_i))\)
    \EndIf
\EndFor
\State \Return \textsc{Alive} list \(\to\) candidate live-variance columns (final eligibility for the benchmark vector additionally requires the semantic / cardinality / sampling checks described in Section~\ref{app-precheck}); \textsc{Dormant} list \(\to\) columns excluded from injection targets and from the vector.
\end{algorithmic}
\end{algorithm}

The procedure of Algorithm~\ref{alg:variance-inventory} requires a single full scan on production-scale tables (\({\sim}1\)--$3$ minutes for 137M rows). It operates on scalar numeric columns only; array and string columns require separate handling (\S\ref{app-precheck}).

\begin{algorithm}[!ht]
\caption{Flip rate verification}\label{alg:flip-rate-check}
\begin{algorithmic}[1]
\Require Injected benchmark table \(T_{\text{inj}}\) containing the labels \(y'\) and \(y_A\) and the per-row drift intensity \(\alpha\); scenario parameters \((\phi, \sigma, \delta)\).
\Ensure Per-scenario empirical flip rate; comparison against schedule-specific theoretical prediction (Eq.~\ref{eq:scenario-flip-rate}).
\ForAll{\(s \in \textsc{Scenarios}\)}
    \State \(n_{\text{total}} \gets \bigl|\{r \in T_{\text{inj}} : \texttt{scenario\_id}(r) = s\}\bigr|\)
    \State \(n_{\text{flipped}} \gets \bigl|\{r \in T_{\text{inj}} : \texttt{scenario\_id}(r) = s \wedge y'(r) \neq y_A(r)\}\bigr|\)
    \State \(\text{flip\_rate}_{\text{empirical}} \gets n_{\text{flipped}} / n_{\text{total}}\)
    \State \(\bar\alpha_s \gets \mathrm{mean}(\alpha)\) over rows with \(\texttt{scenario\_id} = s\) \Comment{schedule-specific}
    \State \(\text{flip\_rate}_{\text{theoretical}} \gets \bar\alpha_s \cdot \phi \cdot \sigma \cdot \delta\) \Comment{Eq.~\ref{eq:scenario-flip-rate}}
    \If{\(\bigl|\text{flip\_rate}_{\text{empirical}} - \text{flip\_rate}_{\text{theoretical}}\bigr| / \text{flip\_rate}_{\text{theoretical}} > 0.05\)}
        \State Warning: scenario \(s\) deviates from design
    \EndIf
\EndFor
\end{algorithmic}
\end{algorithm}

\begin{algorithm}[!ht]
\caption{Ground-truth per-bucket aggregation}\label{alg:ground-truth-check}
\begin{algorithmic}[1]
\Require Injected table \(T_{\text{inj}}\) with per-row ground truth \(m\) (Eq.~\ref{eq:expected-magnitude}) and \(f\) (Eq.~\ref{eq:realized-flip}); number of position buckets \(N\).
\Ensure Per-bucket true expected drift magnitude (primary correlation target) and per-bucket realized-flip rate (robustness target).
\State Define bucket boundaries according to the stream order for \(N\) buckets.
\ForAll{bucket \(b = 0, \ldots, N-1\)}
    \State \(\bar m_b \gets \mathrm{mean}(m)\) over rows in bucket \(b\)
    \State \(\bar f_b \gets \mathrm{mean}(f)\) over rows in bucket \(b\)
\EndFor
\State \Return \((\bar m_b)_{b=0}^{N-1}\) and \((\bar f_b)_{b=0}^{N-1}\): the former is \(\approx 0\) for pre-drift buckets and \(\alpha \cdot \sigma \cdot \phi \cdot \delta\) for buckets at drift intensity \(\alpha\) (where \(\phi\) is scope fraction and \(\delta\) the concept disagreement rate); at production-scale bucket sizes the two arrays agree to \(\Delta r < 10^{-3}\) when used as correlation targets.
\end{algorithmic}
\end{algorithm}

The row-level quantities \(m\) and \(f\) are written by the injection query according to Eqs.~(\ref{eq:expected-magnitude}) and (\ref{eq:realized-flip}).

\begin{algorithm}[!ht]
\caption{Zero-flip check in the pre-drift region}\label{alg:zero-flip-check}
\begin{algorithmic}[1]
\Require Injected table \(T_{\text{inj}}\).
\Ensure Boolean --- has determinism preservation succeeded?
\ForAll{\(s \in \textsc{Scenarios}\)}
    \State \(n_{\text{wrong\_pre}} \gets \bigl|\{r \text{ in scenario } s : \alpha(r) = 0 \wedge y'(r) \neq y_A(r)\}\bigr|\)
    \If{\(n_{\text{wrong\_pre}} > 0\)}
        \State \Return \textsc{Fail} \((s, n_{\text{wrong\_pre}})\)
    \EndIf
\EndFor
\State \Return \textsc{Pass}
\end{algorithmic}
\end{algorithm}

Rationale: In the pre-drift region, \(p_{\text{active}} = p_A\). Since Concept A is a deterministic rule (\(\mathrm{sim}_{\mathrm{img}} > 0\) with integer output 0 or 1), the \(u_{\text{label}} < p_{\text{active}}\) comparison must be deterministic, and \(y' = y_A\) must hold.

\begin{algorithm}[!ht]
\caption{Covariate side-effect distribution check}\label{alg:covariate-stationarity}
\begin{algorithmic}[1]
\Require Injected table \(T_{\text{inj}}\); list of \(X\) columns to check \(C_X\).
\Ensure Per-column pre-drift vs post-drift distribution comparison.
\ForAll{\(c \in C_X\)}
    \State \(\text{pre\_stats} \gets (\mathrm{mean}(c), \mathrm{sd}(c), \mathrm{distinct}(c))\) over rows where \(\alpha(r) = 0\)
    \State \(\text{post\_stats} \gets (\mathrm{mean}(c), \mathrm{sd}(c), \mathrm{distinct}(c))\) over rows where \(\alpha(r) = 1\)
    \If{\(\text{relative\_diff}(\text{pre\_stats}, \text{post\_stats}) > 0.02\)}
        \State Warning: column \(c\) shows aggregate distributional discrepancy from the pre-drift baseline (may indicate that injection has an unintended side effect on this column, or natural production heterogeneity across stream positions)
    \EndIf
\EndFor
\State \Return Provides no evidence of an aggregate distributional side effect on the checked \(X\) summaries (this is a distribution-level consistency check and does not by itself establish row-wise equality \(X_{\text{after}}(r) = X_{\text{before}}(r)\)).
\end{algorithmic}
\end{algorithm}

\section{Position-25 Anomaly: Source-table Heterogeneity Diagnosis}\label{secA4}

Figure~\ref{fig:trendyol-trajectory-rff-swd} reveals a recurring pattern: \emph{all eight scenarios show a systematic spike in the RFF-MMD statistic at position 25}, reaching roughly 20\(\times\) the pre-drift mean. This position sits within the region where \(\bar m_b = 0\); it might be interpreted as a false positive, but the sanity checks defined in Section~\ref{app-sanity} confirm that no row in this position was affected by the injection. To diagnose the behavior, we compared the content-feature profile of the bucket-25 rows against the reference distribution and against the neighboring buckets 24 and 26.

The finding: \emph{position 25 departs systematically from the reference across every measured feature --- this is not an injection artifact but a real data-heterogeneity phenomenon in the source table}. The bucket-25 rows exhibit \(19\%\) higher mean collection size (23.4 vs 19.6) and \(17\%\) higher std (16.9 vs 14.5), $11$--\(13\%\) lower similarity metrics across all three cosine channels (image, text, tag), and \(13\%\) lower mean \(y'\) (0.191 vs 0.219); neighboring buckets 24 and 26 depart by \({<}2\%\) on every feature. This indicates that collections whose collection-identifier hash falls into this bucket carry a systematically different feature profile --- most likely a batch collection deposit within a specific time window, or a taxonomy segment whose hash range happens to align with this bucket --- a one-off, non-periodic anomaly.

This finding highlights three points. First, \emph{RFF-MMD producing a high statistic in this bucket is correct behavior} --- not a ``false positive'' but a correct capture of a real distributional difference in the source data. Because the injection is not active in this region, this reduces the test's correlation against the ground truth (since \(\bar m_b = 0\) enters the correlation) but demonstrates the test's fundamental sensitivity. Second, \emph{this is a real challenge that drift detection faces in production environments}: injected synthetic drift and natural data heterogeneity mix within the same statistical signal. Third, how to handle such anomalies in production monitoring --- requiring an injection-agnostic yet content-aware framework --- is an open research question, noted in Section~\ref{sec8} as future work.

\section{Extension Methods: Full Characterization Results}\label{secA5}

The two extension methods introduced in Section~\ref{subsec4-1} --- RFF-MMD and Sliced Wasserstein Distance --- were characterized under the Failing Loudly protocol; their motivation is to avoid MMD-RBF's \(O(N^2 d)\) scaling. This section reports how successful each method is and which calibration challenges they bring.

\subsection{No-shift calibration of the extension methods}\label{app-subsubsec6-4-1}

Under 60 randomized no-shift comparisons (30 per dataset), an observed FPR whose Wilson 95\% interval contains $0.05$ is compatible with the nominal level in these finite controls (finite-sample chance variation can produce intervals that miss the nominal value even for a well-calibrated test). RFF-MMD satisfies this expectation at all sample sizes, backed by the population-level super-uniformity of the permutation construction (under the exchangeability assumptions of the permutation null). Sliced Wasserstein Distance does not: with the fixed threshold \(\tau_{\mathrm{SW}} = 0.05\), the observed FPR fluctuates between 50\% and 100\% and the Wilson interval never contains $0.05$ at any \(N\) --- the present SWD implementation is not a size-controlled test. At Trendyol production scale (Section~\ref{subsec6-3}), the SWD statistic magnitude does track drift in the strong regime (mean \(r = 0.819\), TPR \(= 82.7\%\)) but under the same fixed threshold the FPR sits around 10\%, consistent with the Failing Loudly calibration finding; a null-percentile calibration analogous to the RFF-MMD procedure (Section~\ref{subsubsec6-3-2}) may improve empirical calibration on similar production distributions and is discussed further in Section~\ref{sec7}.

\subsection{Scalability-accuracy trade-off}\label{app-subsubsec6-4-2}

RFF-MMD (\(D = 500\)) has \(O(N \cdot D \cdot d)\) compute complexity, fundamentally different from MMD-RBF's \(O(N^2 d)\), and scales to \(N = 10\,000\) and beyond (the Trendyol scans apply it with \(D = 200\) against a 100,000-row reference). Under Failing Loudly, RFF-MMD detection accuracy remains comparable to MMD-RBF (98\% on MNIST combined shifts, 15\% on CIFAR-10 adversarial shifts vs.\ 7\% for MMD-RBF) with reliable permutation size control, and in the Trendyol benchmark (Section~\ref{subsec6-3}) produced the most consistent strong-regime detection performance (mean \(r = 0.940\) across four drift types on \({\sim}137.5\)M rows).

\section{Synthetic Label Design: Concept A and Concept B Definitions}\label{secA6}

Since the source table contains no genuine target column that could serve as a drift injection substrate, we defined Concept A and Concept B rules over the live similarity metrics.

Concept A labels a row positive if the collection shows any visual similarity to its seed content:
\begin{equation}\label{eq:concept-a-rule}
    Y_A = \indicator[\mathrm{sim}_{\mathrm{img}} > 0],
\end{equation}
where \(\mathrm{sim}_{\mathrm{img}}\) is the image cosine similarity between the collection and its seed content. This similarity is bimodal in the source table --- 78\% of rows are zero and the remaining 22\% lie in \([0.5, 1.1]\) (median 0.94), with almost no intermediate values --- so this rule functions as a near-binary ``was similarity measured'' signal with positive rate \(P(Y_A = 1) \approx 0.2195\).

Concept B labels a row positive if there is any textual similarity:
\begin{equation}\label{eq:concept-b-rule}
    Y_B = \indicator[\mathrm{sim}_{\mathrm{txt}} > 0],
\end{equation}
where \(\mathrm{sim}_{\mathrm{txt}}\) is the corresponding text cosine similarity, with \(P(Y_B = 1) \approx 0.4473\). The \emph{disagreement rate} between the two rules --- the fraction of rows where they yield different decisions --- is \(\delta \approx 0.5724\) over the source table, providing an upper bound on the maximum theoretical drift effect (no more than a \(\delta\)-fraction of rows can have their labels changed even at maximum severity and scope).

This concept pair was chosen for three reasons: (i) \emph{conceptual separation} --- visual vs.\ textual similarity are distinct signal modalities of the same recommendation domain, mirroring a realistic shift from vision-based to text-based signals in a production ranking pipeline; (ii) \emph{balanced baselines} --- both positive rates fall in \([0.22, 0.45]\), neither too sparse nor too dominant for Bernoulli-based two-sample tests; and (iii) \emph{derivation from \(X\)} --- \(Y\) is deterministically derived from features present in the table, so the injection can be interpreted as a change in \(P(Y \mid X)\).

\section{Injection Boundaries: The Covariate-Preservation Guarantee}\label{secA7}

A critical design principle of the pipeline is that no feature in the source table is modified by injection: the injected label \(y'\) is added as an outcome, but the similarity metrics (including \(\mathrm{sim}_{\mathrm{img}}\) and \(\mathrm{sim}_{\mathrm{txt}}\)), the recommendation source, the collection size, and all other original columns remain untouched. This guarantees the injection operator leaves \(X\) row-wise unchanged (\(X_{\text{after}}(r) = X_{\text{before}}(r)\) for every row) and provides the analytical ground on which the injection-relative component of the signal can be attributed to the change in \(P(Y \mid X)\). Because the underlying production table may carry natural covariate heterogeneity across stream positions, the observed reference and current window distributions need not satisfy \(P_R(X) = P_C(X)\) exactly (empirically visible at position 25; Section~\ref{subsubsec6-3-6}); any such departure is a source-table phenomenon rather than a pipeline artifact.

Algorithm~\ref{alg:concept-drift-injection} formalizes the row-level injection process defined in Section~\ref{subsec5-4}, expressed in terms of a table \(\mathcal{T}\) with rows indexed by the stream index \(i\) and the parameters \((\sigma, \phi, \text{drift type})\) defining a scenario. Executable as a single-pass SQL query or DataFrame transformation over any distributed backend (BigQuery, Spark, pandas).

\begin{algorithm}[!ht]
\caption{Concept Drift Injection}\label{alg:concept-drift-injection}
\begin{algorithmic}[1]
\Require Source table \(T\) with \(n\) rows; drift type \(d \in \{\textsf{sudden}, \textsf{gradual}, \textsf{incremental}, \textsf{recurring}\}\); severity \(\sigma \in [0,1]\); scope \(\phi \in [0,1]\); drift start fraction \(s_0\), drift width fraction \(w\), recurrence period fraction \(\lambda\); random seed \(\zeta\); hash modulus \(M\); Concept A rule \(Y_A\), Concept B rule \(Y_B\) over \(X\).
\Ensure Drifted table \(T'\) with synthetic target \(y'\), drift intensity \(\alpha\), and per-row ground truth \(e(r)\), \(m(r)\), \(f(r)\) (Eqs.~\ref{eq:eligible}--\ref{eq:realized-flip}); the coarser indicator \(a(r) = \indicator[\alpha > 0 \wedge e(r) = 1]\) is also emitted for backward comparability.
\State \(n_s \gets \lfloor n \, s_0 \rfloor\); \; \(n_w \gets \max(1,\, \lfloor n \, w \rfloor)\); \; \(n_p \gets \max(1,\, \lfloor n \, \lambda \rfloor)\)
\ForAll{\(r \in T\) in ascending order of \(\mathrm{hash}(\mathrm{id}(r))\), with rank \(i = 0, 1, \ldots, n-1\)}
    \If{\(d = \textsf{sudden}\)} \State \(\alpha \gets 0\) if \(i < n_s\) else $1$
    \ElsIf{\(d = \textsf{incremental}\)} \State \(\alpha \gets \mathrm{clip}\bigl((i - n_s)/n_w,\, 0,\, 1\bigr)\)
    \ElsIf{\(d = \textsf{gradual}\)}
        \If{\(i < n_s\)} \State \(\alpha \gets 0\)
        \ElsIf{\(i \geq n_s + n_w\)} \State \(\alpha \gets 1\)
        \Else{} \State \(\alpha \gets \bigl[1 + \exp\bigl(-12 \bigl((i-n_s)/n_w - 0.5\bigr)\bigr)\bigr]^{-1}\)
        \EndIf
    \ElsIf{\(d = \textsf{recurring}\)}
        \State \(\alpha \gets 0\) if \(i < n_s\) else \(\bigl(\lfloor (i - n_s)/n_p \rfloor \bmod 2\bigr)\)
    \EndIf
    \State \(u_{\mathrm{scope}} \gets \mathrm{hash}\bigl(\mathrm{id}(r),\, \zeta,\, \text{``scope''}\bigr) / M\)
    \State \(\text{in\_scope} \gets \indicator[u_{\mathrm{scope}} < \phi]\)
    \State \(p_A \gets Y_A(X(r))\); \; \(p_B \gets Y_B(X(r))\) \Comment{deterministic rules over \(X\)}
    \If{\(\text{in\_scope}\)}
        \State \(p_B^{\mathrm{eff}} \gets p_A + \sigma \cdot (p_B - p_A)\) \Comment{severity-scaled Concept B, Eq.~(\ref{eq:active-target-prob})}
    \Else
        \State \(p_B^{\mathrm{eff}} \gets p_A\)
    \EndIf
    \State \(p_{\mathrm{active}} \gets p_A + \alpha \cdot (p_B^{\mathrm{eff}} - p_A)\) \Comment{Eq.~(\ref{eq:active-target-prob})}
    \State \(u_{\mathrm{label}} \gets \mathrm{hash}\bigl(\mathrm{id}(r),\, \zeta,\, \text{``label''}\bigr) / M\)
    \State \(y'(r) \gets \indicator\bigl[u_{\mathrm{label}} < p_{\mathrm{active}}\bigr]\)
    \State \(e(r) \gets \indicator[\text{in\_scope} \wedge p_A \neq p_B]\) \Comment{Eq.~(\ref{eq:eligible})}
    \State \(m(r) \gets \alpha \cdot \sigma \cdot e(r)\) \Comment{Eq.~(\ref{eq:expected-magnitude})}
    \State \(f(r) \gets \indicator[y'(r) \neq y_A(r)]\) \Comment{Eq.~(\ref{eq:realized-flip})}
    \State \(a(r) \gets \indicator[\alpha > 0 \wedge e(r) = 1]\) \Comment{coarser indicator retained for backward comparability}
    \State \textbf{emit} \(\bigl(r,\ y'(r),\ \alpha,\ e(r),\ m(r),\ f(r),\ a(r)\bigr)\)
\EndFor
\end{algorithmic}
\end{algorithm}

\end{appendices}


\end{document}